\documentclass[preprint,review,12pt]{elsarticle}

\usepackage{amssymb} %%希腊字母
\usepackage{amsmath} %%数学公式
\usepackage[ruled]{algorithm2e} %%算法
\usepackage{colortbl,booktabs} %%表格
\usepackage{tabularx} %%自动计算列宽
\usepackage{threeparttable} %%给表格添加注释
\usepackage{multicol,multirow} %%绘制表格
\usepackage{graphicx}
\usepackage{subfigure} 
\usepackage{enumerate}
\usepackage [font=footnotesize,labelfont=bf]{caption} %%控制题目/说明文字的样式
\usepackage{rotating} %%旋转文字,图像,表格等对象
\usepackage{tikz} %%绘图
\usepackage[colorlinks,linkcolor=blue]{hyperref}
\usepackage{color}
\definecolor{R}{rgb}{0.9, 0.0, 0.0}
\definecolor{G}{rgb}{0.0, 0.9, 0.0}
\definecolor{B}{rgb}{0.0, 0.0, 0.9}

\journal{Elsevier}

\begin{document}
\begin{frontmatter}
	\title{Target-driven splitting SPH optimization of 
		   thermal conductivity distribution}
	\author[author1]{Bo Zhang}
	\ead{bo.zhang.aer@tum.de}
	\author[author2]{Chi Zhang}
	\ead{c.zhang@tum.de}
	\author[author1]{Xiangyu Hu\corref{mycorrespondingauthor}}
	\cortext[mycorrespondingauthor]{Corresponding author.}
	\ead{xiangyu.hu@tum.de}
	\address[author1]{TUM School of Engineering and Design,
		              Technical University of Munich,\\
		              85748 Garching, Germany}
	\address[author2]{Huawei Technologies Munich Research Center \\
	                  80992 Munich, Germany}
	
	\begin{abstract}
		Efficiently enhancing heat conduction through optimized 
		distribution of a limited quantity of high 
		thermal conductivity material is paramount in cooling 
		electronic devices and numerous other applications.
		This paper introduces a target-driven all-at-once approach 
		for PDE-constrained optimization and 
		derives a splitting smoothed particle hydrodynamics (SPH) 
		method for optimizing the distribution of 
		thermal conductivity in heat conduction problems. 
		In this method, the optimization iteration of the system 
		is split into several easily addressed steps.
		A targeting step is employed to progressively enforce the 
		direct target, which potentially leads to increased PDE residuals.
		Then, these residuals are recovered 
		through an evolution step of the design variable.
		After this, a PDE solution step is carried out 
		to further decrease the PDE residuals, 
		and the system is ready for the next iteration. 
		Unlike the simulation-based approaches, 
		the present method does not rely on the
		adjoint state equation and converged state variable field
		in each iteration, and the optimization process 
		is significantly simplified and accelerated.
		With the utilization of an implicit SPH splitting operator 
		and a general numerical regularization formulation, 
		the information propagation is further accelerated 
		and the numerical stability is greatly enhanced.
		Typical examples of heat conduction optimization demonstrate 
		that the current method yields optimal results comparable 
		to previous methods and exhibits considerable computational efficiency.
		Moreover, the optimal results feature 
		more moderate extreme values,
		which offers distinct advantages for the easier selection of 
		appropriate material with high thermal conductivity.
	\end{abstract}
	
\end{frontmatter}

% \linenumbers
%%%%%%%%%%%%%%%%%%%%%%%%%%%%%%%%%%%%%%%%%%%%%%%%%%%%%%%%%%%%%
%
% Section                                                                                           
%
%%%%%%%%%%%%%%%%%%%%%%%%%%%%%%%%%%%%%%%%%%%%%%%%%%%%%%%%%%%%%
\section{Introduction}\label{introduction}
Electronic devices find wide-ranging applications in 
various industries, including aerospace, transportation, 
network communication, processing, and manufacturing. 
The evolution of microelectronics has facilitated the 
miniaturization of electronic devices, but this reduction 
in device size drastically increases power density
and presents considerable challenges for effective cooling \cite{moore2014emerging}.
To guarantee the expected performance and lifespan, 
multiple design ideas have been explored to dissipate the 
generated heat and lower the operating temperatures
\cite{almogbel1999conduction, guo2003least, 
	da2005distribution, chen2016optimization,
	feng2015constructalflowpassage, ghani2017hydrothermal}.
Among these ideas, inserting high thermal conductivity material 
\cite{almogbel1999conduction, guo2003least} has garnered extensive 
attention due to its effectiveness and simplicity of implementation.
The rapid innovation in additive manufacturing, specifically in metal 
3D printing, has also opened up new possibilities for implementing 
this cooling concept \cite{dede2015topology, zegard2016bridging}.

Thus, the primary design challenge lies in finding the optimal distributions 
of high thermal conductivity materials to meet specific cooling targets. 
This optimization is also known as the volume-to-point (VP) 
problem \cite{bejan1997constructal}.
It involves redistributing a fixed amount (constraint) of material 
with high thermal conductivity (design variable) to cool 
a heat-generating volume within specified boundaries, 
with the objective of minimizing the temperature (state variable) 
in the given domain. 
The VP problem is a typical example of PDE-constrained optimization, 
characterized by the constraint of physical principles expressed as 
partial differential equations (PDEs) \cite{de2015numerical}.

The methods for PDE-constrained optimization can be categorized 
into simulation-based and all-at-once approaches, 
depending on how the PDE constraint is handled 
\cite{herzog2010lectures, herzog2010algorithms}. 
In simulation-based methods, it is eliminated by 
obtaining a converged physical solution using existing solvers.
Therefore, the optimization process computes the gradients of state 
variables with respect to the design ones at the hypersurface of 
the physical solution to determine the searching direction in each iteration.
Typical approaches for computing gradients, whether done explicitly or implicitly,
include the adjoint technique \cite{ito2008lagrange, zhao2022optimal},
automatic differentiation (AD) \cite{rall1996introduction} and 
artificial neural network (ANN) \cite{abiodun2018state, song2021optimization}.
Although the simulation-based methods are conceptually appealing 
and widely used in solving VP problems \cite{chen2011alternative, 
	qi2015assessment, chen2019entropy},
they require repeated and costly solutions of the PDEs, even in the initial 
stages when the design variables are still far from their optimal values.
All-at-once methods, which explicitly maintain the PDE constraint as 
another optimal target and treat both the state and design variables equally.
A clear advantage is that they do not require repeated PDE solutions
but satisfy the PDE constraint only at the termination of optimization.
Some all-at-once methods, 
such as the augmented Lagrangian method \cite{bertsekas2014constrained} 
and sequential quadratic programming method \cite{alt1993lagrange}, 
have been proposed for addressing general PDE-constrained optimization problems. 
However, these methods are rare in solving VP problems 
due to their inherent complexity and high computational costs.

In early works on simulation-based methods for tackling the VP problem, 
some indirect global optimal principles, such as 
temperature gradient field homogenization 
\cite{cheng2003constructs, cheng2009homogenization, guo2007entransy},
entropy generation minimization (EGM)
\cite{bejan1983entropy, bejan1996entropy, du2015optimization, wang2020constructal},
and entransy dissipation extremum (EDE)
\cite{guo2007entransy, chen2013entransyreview, 
	zhao2022irreversibility, zhao2019collaborative, wu2022study},
have been proposed and adopted.
Although these principles were expected to be equivalent to 
the direct target of the lowest average temperature,
it has been argued that they are not equivalent 
and directly utilizing these principles 
for different targets or boundaries may 
lead to sub-optimal or even detrimental solutions 
\cite{zhao2022irreversibility}.
Therefore, direct optimization targets,
including the lowest average \cite{tong2018optimizing, zhao2022optimal}
and minimizing hot spot temperature \cite{song2021optimization,zhang2023effective},
have also been employed recently.

In addition to the aforementioned iterative methods based on 
deterministic principles, some stochastic approaches, such as
the bionic optimization (BO) method \cite{xia2002heat, xia2004bionic},
cellular optimization (CA) method \cite{boichot2009tree, boichot2010simple},
simulated annealing (SA) and genetic algorithm (GA) method \cite{xu2007optimization},
have been explored to address VP problems.
Topology optimization, for example the asymptotes algorithm (MMA) \cite{burger2013three} 
and density-based method \cite{manuel2017design}, 
has also gained attention for optimal heat conduction problems.
While these stochastic methods offer novel avenues for achieving optimal solutions, 
they often deal with a large number of variables in the spatial discretization 
of the domain, leading to reduced efficiency and suitability for large-scale 
and reliability-sensitive problems.
Therefore, developing an efficient yet straightforward method, particularly 
leveraging the direct optimization target and the all-at-once concept,
holds great promise for addressing heat conduction optimization problems.

With these in mind, this paper introduces a target-driven 
all-at-once method for PDE-constrained optimization problems 
and derives a smoothed particle hydrodynamics (SPH) method 
for optimizing the thermal conductivity 
distribution to minimize the average temperature.
In this method, the optimization iteration 
is divided into several easily handled steps.
A targeting step is adopted ot progressively impose the direct target,
potentially resulting in increased PDE residuals.
Then, through an evolution step of the design variables, 
these residuals are subsequently recovered.
Following this, a PDE solution step is performed 
to further decrease the PDE residual and prepare 
for the next iteration.
The novelty of this work can be summarized in three aspects.
Firstly, as the split steps are only weakly coupled 
with each other, compared to previous all-at-once approaches, 
the present updating of both state and design variables is 
much easier to handle.
Secondly, by leveraging the splitting-operator SPH method,
implicit updating is achieved without the inversion of 
large-size matrices.
Thirdly, a general formulation of regularization 
has been proposed to achieve numerical stability 
when evolving the design variable.

In the following sections, 
Section \ref{problemdescription} provides a brief problem description;
Section \ref{numericalscheme} introduces the SPH method and the numerical scheme; 
Section \ref{targetdrivenoptimization} elucidates the proposed 
optimization method, 
and Section \ref{results} demonstrates typical examples to 
validate the effectiveness of the method.
Finally, Section \ref{conclusion} concludes the key findings and 
outlooks the future work.
%
%%%%%%%%%%%%%%%%%%%%%%%%%%%%%%%%%%%%%%%%%%%%%%%%%%%%%%%%%%%%%
%
% Section
%
%%%%%%%%%%%%%%%%%%%%%%%%%%%%%%%%%%%%%%%%%%%%%%%%%%%%%%%%%%%%%
\section{Problem description}\label{problemdescription}
\begin{figure}[htb!]
	\centering
	\includegraphics[width=0.8\textwidth]{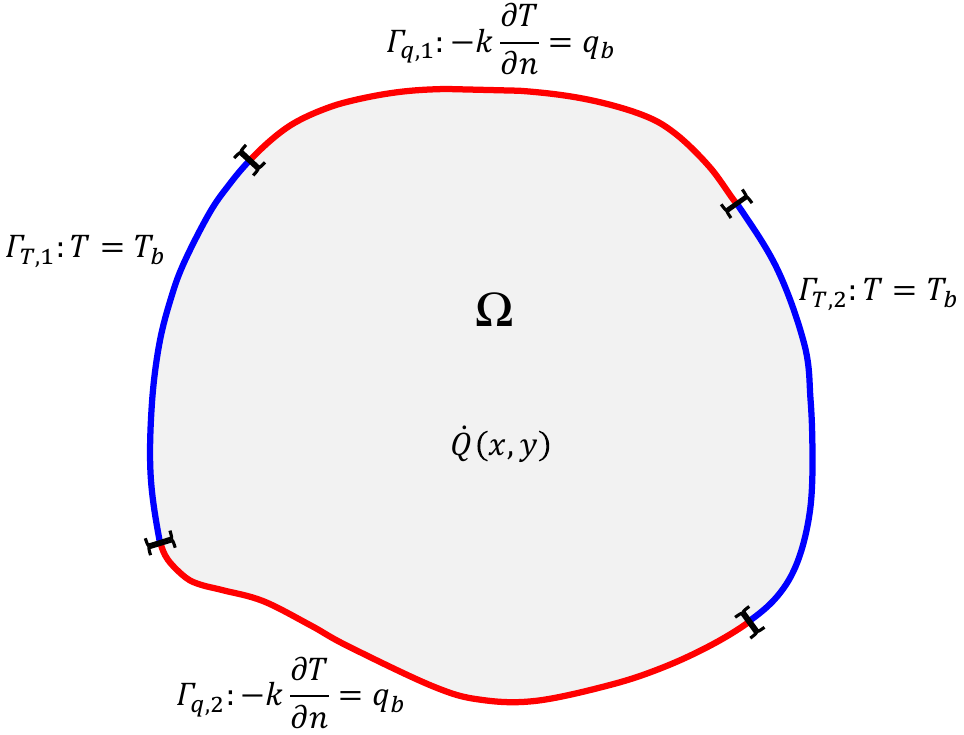}
	\caption{Illustration of the 2D heat conduction problem.
		$\Omega$ denotes the thermal domain;
		$\varGamma_{T}$ and $\varGamma_{q}$ represent constant 
		surface temperature and heat flux boundaries, respectively,
		and $\dot{Q}$ indicates the presence of an internal heat source.}
	\label{thermaldomainsketch}
\end{figure}

%%%%%%%%%%%%%%%%%%%%%%%%%%%%%%%%%%%%%%%%%%%%%%%%%%%%%%%%%%%%%
We consider the optimization of thermal conductivity 
distribution for two-dimensional heat conduction problems,
specifically addressing typical VP problems.
As illustrated in Fig. \ref{thermaldomainsketch},
the thermal domain under consideration, denoted as $\Omega$, 
is subject to different boundary conditions and may contain internal heat sources.
The steady-state temperature field within the thermal domain 
can be determined by solving the transient heat conduction governing equation 
given by
\begin{equation}
	\dfrac{\text{d}T}{\text{d}t} = \nabla\cdot\left(k\nabla T\right)
	+ \dot{Q} \quad \text{in} \quad \Omega,
	\label{heatconductiongoverningequation}
\end{equation}
when the left-hand side (LHS) converges to zero.
Here, $T$ represents temperature, 
$k$ denotes thermal conductivity, 
and $\dot{Q}$ is the volume rate of the internal heat source.
Note that, for the sake of simplified analysis,
Eq. \eqref{heatconductiongoverningequation} is reduced 
from the general heat conduction equation with $\rho C = 1$, 
where $\rho$ and $C$ represent density and heat capacity, respectively.
	
Two typical boundary conditions are considered here, 
as shown in Fig. \ref{thermaldomainsketch}. 
The Dirichlet boundary condition is given as
\begin{equation}
	T = T_b \quad \text{on} \quad \varGamma_{T},
	\label{DirichletBoundaryCondition}
\end{equation}
where the surface temperature $T_b$ is held constant.
The Neumann boundary condition, 
which maintains a fixed heat flux rate, can be expressed as,
\begin{equation}
	-k\nabla T\cdot \boldsymbol{n} = q_b \quad \text{on} \quad \varGamma_{q},
	\label{NeumannBoundaryCondition}
\end{equation}
where $\boldsymbol{n}$ indicates the surface normal vector pointing outward.

The objective of the optimization is to obtain 
a distribution of thermal conductivity in $\Omega$
by minimizing the average steady temperature 
$\overline{T}$ defined as
\begin{equation}
	\overline{T} = \frac{1}{V}\int_{V} T dV,
	\label{objectivefunction}
\end{equation}
where $V$ is the total volume of the thermal domain.
Additionally, the average thermal conductivity is constrained 
to remain constant by
\begin{equation}
	\int_{V} k dV  =k_0 V, 
	\label{thermalconductivityconstrain}
\end{equation}
where $k_0$ is a reference value throughout the optimization process.
%
%%%%%%%%%%%%%%%%%%%%%%%%%%%%%%%%%%%%%%%%%%%%%%%%%%%%%%%%%%%%%
%
% Section
%
%%%%%%%%%%%%%%%%%%%%%%%%%%%%%%%%%%%%%%%%%%%%%%%%%%%%%%%%%%%%%
\section{Numerical scheme}\label{numericalscheme}
%%%%%%%%%%%%%%%%%%%%%%%%%%%%%%%%%%%%%%%%%%%%%%%%%%%%%%%%%%%%%
\subsection{SPH formulation}\label{SPH}
In this study, the SPH method is employed to solve the 
temperature and thermal conductivity fields.
SPH is a fully Lagrangian particle method that was 
initially proposed for astrophysical applications 
\cite{lucy1977numerical, gingold1977smoothed}. 
Since its introduction, SPH has demonstrated significant 
success in simulating a wide range of scientific problems,
including heat transfer problems 
\cite{vishwakarma2011steady, xiao2017modeling}.

In the SPH scheme, the heat conduction governing equation 
in Eq. \eqref{heatconductiongoverningequation} 
can be discretized at each SPH particle $i$ located at $\boldsymbol{r}_i$
with its neighboring particles $j$ as follows,
\begin{equation}
	\displaystyle\dfrac{\text{d}T_i}{\text{d}t}=2 \sum_{j}
	\overline k_{ij}
	\displaystyle\dfrac{T_{ij}}{r_{ij}} \nabla_{i} W_{ij} V_j + \dot{Q}_{i}.
	\label{heatconductionsphformulation}
\end{equation}
Here, $\nabla_{i} W_{ij}= \nabla_{i} W\left(\left|\boldsymbol{r}_{ij}\right|,h\right)
= \frac{\partial W_{ij}}{\partial r_{ij}} \boldsymbol{e}_{ij}$,
where $\boldsymbol{r}_{ij} = \boldsymbol{r}_i - \boldsymbol{r}_j$,
$h$ is the smoothing length 
and the unit vector $\boldsymbol{e}_{ij} = \frac{\boldsymbol{r}_{ij}}{r_{ij}}$,
represents the derivative of the kernel function.
$T_{ij} = T_i -T_j$ indicates the inter-particle temperature difference,
and $V_{j}$ denotes the volume of neighboring particles $j$.
The quantity $\overline k_{ij} = 
(k_i+k_j)/2$ is the inter-particle average thermal conductivity.

Near the domain boundary, several layers of dummy particles are introduced 
to enforce different boundary conditions.
Implementing the Dirichlet boundary condition is straightforward 
and involves imposing the temperatures 
\begin{equation}
	T_w = 2 T_b - T_i,
\end{equation}
at dummy particles 
implied by the wall boundary condition \cite{adami2012generalized}.
To implement the Neumann boundary condition, 
the discretization of Eq. \eqref{heatconductiongoverningequation} 
is modified into
\begin{equation}
	\dfrac{\text{d}T}{\text{d}t} = \nabla\cdot \left(k\nabla T\right)
	+ \dot{Q} + \dot{Q}^{\varGamma q} \quad \text{in} \quad \Omega,
	\label{CBFmethodforNeumannBoundaryCondition}
\end{equation}
following the Ref. \cite{ryan2010novel},
where the heat flux in Eq. \eqref{NeumannBoundaryCondition}
is replaced by a volumetric term $\dot{Q}^{\varGamma q}$,
which can be discretized as
\begin{equation}
	\dot{Q}^{\varGamma q}_{i} = -q_b \displaystyle\sum_{j\in\Omega_{j}} 
	\left(\boldsymbol{n}_{i}+\boldsymbol{n}_{j}\right) \cdot \nabla_{i} W_{ij}V_j.
\end{equation}
Here, $\Omega_{j}$ represents the boundary domain defined by the dummy particles.
The unit vectors $\boldsymbol{n}_i$ and $\boldsymbol{n}_j$ are normal 
to the boundary evaluated at the respective positions of particle $i$ and $j$.

\subsection{Splitting operator based implicit scheme}\label{implicit}
It is well known that traditional implicit schemes often require 
large-scale matrix inversion or iterations across the entire system, which 
can lead to significant memory demands and challenges in parallelization.
To overcome these challenges, 
we employ a splitting operator based implicit scheme
to advance Eq. \eqref{heatconductionsphformulation}.
The implicit solving step is divided into individual 
particle-by-particle operations, 
and each evolves a small system that is easy to inverse.
One commonly used approach for this purpose is the second-order
Strang splitting technique \cite{strang1968construction}, shown as
\begin{equation}
	\begin{aligned}
		S_i^{(\Delta t)} = &D_{1}^{(\frac{\Delta t}{2})}  
		\circ D_{2}^{(\frac{\Delta t}{2})} \circ \dots 
		\circ D_{i}^{(\frac{\Delta t}{2})} \dots 
		\circ D_{N_t-1}^{(\frac{\Delta t}{2})} 
		\circ D_{N_t}^{(\frac{\Delta t}{2})} \circ \\ 
		&D_{N_t}^{(\frac{\Delta t}{2})}  
		\circ D_{N_t-1}^{(\frac{\Delta t}{2})} \circ \dots 
		\circ D_{i}^{(\frac{\Delta t}{2})} \dots 
		\circ D_{2}^{(\frac{\Delta t}{2})} 
		\circ D_{1}^{(\frac{\Delta t}{2})}.
	\end{aligned}
	\label{particlebyparticlesplitting}
\end{equation}
Here, the operator $S_i^{(\Delta t)}$ represents the 
complete step for advancing the equation.
$N_t$ refers to the total number of particles, and $D_i$ 
represents the splitting operator corresponding to particle $i$. 
The update of the variable for the entire field involves 
a forward sweep of all particles for half a time step, followed 
by a backward sweep for another half time step \cite{nguyen2009mass}.

Within the local implicit formulation, 
Eq. \eqref{heatconductionsphformulation} can be rewritten as
\begin{equation}
	\displaystyle\dfrac{\text{d}T_{i}}{\text{d}t}=2 \sum_{j} \overline k_{ij} 
	\displaystyle\dfrac{T_{ij}^{n+1}}{r_{ij}} 
	\nabla_{i} W_{ij} V_j + \dot{Q}^{n+1}_{i},
	\label{heatconductionimplicitformulation}
\end{equation}
where $T_{ij}^{n+1} = T_{ij}^{n} +\text{d}T_i-\text{d}T_j$.
The terms $\text{d}T_i$ and $\text{d}T_j$ represent 
the incremental changes for particle $i$ and its neighboring 
particles $j$ at each advancing time step.
For brevity, we introduce the coefficient 
\begin{equation} 
	B_j = 2\overline{k}_{ij}\displaystyle\dfrac{1}{r_{ij}} 
	\nabla_{i}W_{ij} V_j \text{d}t,
	\label{implicitcoefficient}
\end{equation}
and the residual of Eq. \eqref{heatconductionimplicitformulation},
without considering the increment, has the form
\begin{equation}
	E_i= -\displaystyle\sum_{j} B_{j}T_{ij}^{n} -\dot{Q}^{n+1}_{i} \text{d}t.
	\label{implicitresidual}
\end{equation}
The implicit formulation of Eq. \eqref{heatconductionimplicitformulation}
can be further expressed as 
\begin{equation}
	E_i = \left(\displaystyle\sum_{j} B_j - 1\right)\text{d}T_i - 
	\displaystyle\sum_{j} B_j \text{d}T_j.
	\label{implicitfurtherresidual}
\end{equation}
To determine the incremental changes for temperature, 
we employ the gradient descent method \cite{bishop2006pattern} by reducing 
the LHS of Eq. \eqref{implicitresidual} following its gradient.
The gradient $\nabla E_i$ with respect to the variable 
$\left(\text{d}T_i, \text{d}T_1, \text{d}T_2, \cdots, \text{d}T_N\right)^{T}$, 
where $N$ gives total number of all neighboring particles,
can be obtained as
\begin{equation}
	\nabla E_{i} = \left(\sum_{j}B_{j}-1, -B_{1}, -B_{2}, 
	\cdots, -B_{N}\right)^{T}.
	\label{gradient}
\end{equation}
We set
\begin{equation}
	\left(\text{d}T_i, \text{d}T_1, \text{d}T_2, \cdots, 
	\text{d}T_N\right)^{T} = \eta_{i}\nabla E_{i},
	\label{learningrate}
\end{equation}
where $\eta_i$ represents the learning rate \cite{bishop2006pattern}
for the particle $i$.
Substituting Eqs. \eqref{gradient} and \eqref{learningrate}
into Eq. \eqref{implicitfurtherresidual}, 
the learning rate can be obtained as
\begin{equation}
	\eta_{i} = \left(\left(\sum_{j}B_j-1\right)^2+\sum_{j}
	\left(B_j\right)^2\right)^{-1}E_i.	
\end{equation}
According to Eqs. \eqref{gradient} and \eqref{learningrate},
the incremental change in temperature of particle $i$ and all 
its neighbors can be obtained and updated as
\begin{equation} \label{splitting-updating}
	\begin{cases}
		\displaystyle T_i^{n+1}&=T_i^{n} - \text{d} T_i = 
		T_i^{n} + \eta_{i} \left(\sum_{j} B_j - 1\right)\\
		\displaystyle T_1^{n+1}&=T_1^{n} - \text{d}T_1 = 
		T_1^{n}  - \eta_{i} B_1\\
		\displaystyle T_2^{n+1}&=T_2^{n} - \text{d}T_2 = 
		T_2^{n}  - \eta_{i} B_2\\
		&\cdots \\
		\displaystyle T_N^{n+1}&=T_N^{n} - \text{d}T_N = 
		T_N^{n}  - \eta_{i} B_N\\
	\end{cases}.
\end{equation}
Note that, Eq. \eqref{splitting-updating} involves updating 
the variables for particle $i$ and its 
neighboring particles simultaneously.
When a shared-memory parallelization is employed,
conflict may arise when multiple threads attempt to update 
the values of a single particle pair simultaneously.
To address this issue, we have implemented a splitting 
Cell Linked List method \cite{zhu2022dynamic}. 
This method effectively prevents conflicts by ensuring that 
neighboring particles are located in the same cell or 
in adjacent cells that are distributed among the same threads. 
Also note that, for an explicit integration of the thermal diffusion equation, 
the maximum allowable time step can be defined as
\begin{equation}
	\Delta t_d = 0.5\dfrac{\rho C h^2}{k_{max}}.
\end{equation}
Since the implicit scheme is employed here for obtaining the steady 
solution of the Eq. \eqref{heatconductionsphformulation}, the time 
step size is chosen as a large value of $10 \Delta t_d$ without 
considering the temporal accuracy.
%
%%%%%%%%%%%%%%%%%%%%%%%%%%%%%%%%%%%%%%%%%%%%%%%%%%%%%%%%%%%%%
%
% Section
%
%%%%%%%%%%%%%%%%%%%%%%%%%%%%%%%%%%%%%%%%%%%%%%%%%%%%%%%%%%%%%
\section{Target-driven optimization}\label{targetdrivenoptimization}
%%%%%%%%%%%%%%%%%%%%%%%%%%%%%%%%%%%%%%%%%%%%%%%%%%%%%%%%%%%%%
\subsection{Method overview}\label{overview}
The target-driven PDE-constrained optimization method proposed 
here is based on the principle of residual recovery.
The primary objective is to solve the PDEs while progressively 
imposing the target function directly defined by the state variable.
As the latter part can potentially lead to modifications 
or even an increase in the PDE residuals,
to address this issue, the design variable undergoes an evolution to 
recover the original residual, before the PDE solving continues.
This process is repeated iteratively until the fields of both state 
and design variables converge and reach steady states.
At that point, the optimal distribution of the design variable is 
obtained while satisfying both the target function and PDE constraints.
Note that, the residual recovery approach is a typical 
all-at-once method since the residual of PDE only converges 
upon completing the optimization process.
 
\begin{figure}[htbp]
	\centering
	\makebox[\textwidth][c]{\includegraphics[width=1.0\textwidth]{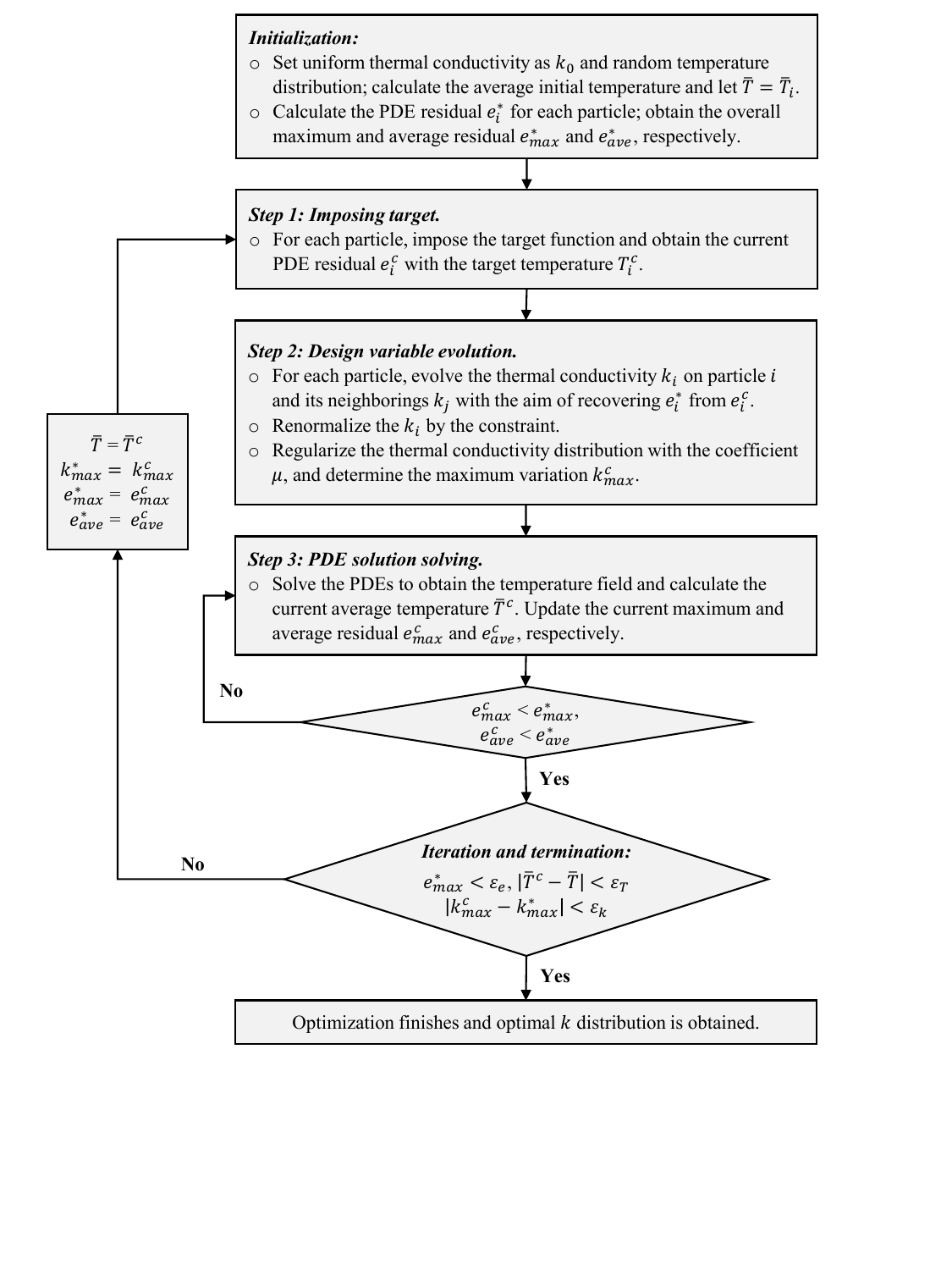}}
	\caption{The flowchart of the target-driven optimization procedure.}
	\label{flowchart}	
\end{figure}

Although the approach proposed above can theoretically be 
applied to general PDE-constrained optimization problems,
here we apply it to address heat-conduction based optimizations
The flowchart of the proposed method is illustrated in Fig. \ref{flowchart}
and the detailed steps are given as follow.
\begin{enumerate}[]
	\item \textbf{Initialization:} 
	\begin{itemize}
		\item 
		The thermal domain is populated with inner and dummy particles 
		for implementing the SPH method.
		The thermal conductivity is initialized as a uniform 
		distribution with $k_0$.
		The initial temperature is randomly assigned, and then, 
		the average initial temperature $\overline{T}$ is calculated.
		\item 
		The PDE residual $e_i^*$,
		i.e. the LHS of Eq. \eqref{heatconductionsphformulation},
		is calculated for each particle within the thermal domain.
		The overall maximum and average residuals, denoted as $e_{max}^{*}$, 
		and $e_{ave}^{*}$, respectively, are then determined.
	\end{itemize}
	\item \textbf{Step 1: Imposing target.}
	\begin{itemize}
		\item The target is locally imposed on each particle 
		with a strength $\beta$. 
		In this case, it can be expressed as $T_{i}^{c} = T_{i}^{*}- 
		\beta$ based on the Eq. \eqref{objectivefunction}.
		The modified PDE residual $e_{i}^{c}$ on each particle is 
		then updated with the imposed temperature $T_{i}^{c}$.
	\end{itemize}
    \item \textbf{Step 2: Design variables evolution.} 
    \begin{itemize}
    	\item The thermal conductivity $k_{i}$ on each particle 
    	and its neighboring particles $k_{j}$ are evolved with the aim 
    	of recovering the residual $e_i^*$ from $e_{i}^{c}$, 
    	as detailed in Section \ref{sectiondesignvariableevolution}.
    	\item Re-normalize the thermal conductivity $k_{i}$ on 
    	each particle to fulfill the constraint presented in 
    	Eq. \eqref{thermalconductivityconstrain}.
    	\item The thermal conductivity distribution is regularized 
    	with the coefficient $\mu$, employing the diffusion analogy 
    	numerical regularization method, as explained in detail in 
    	Section \ref{sectionregularization}.
    	The maximum thermal conductivity variation, denoted as $k^{c}_{max}$, 
    	is determined by Eq. \eqref{regularizationformulation}.
    \end{itemize}
    \item \textbf{Step 3: PDE solution solving.}
    \begin{itemize}
    	\item The PDE solving advances 
    	using the updated thermal conductivity values 
    	to obtain the intermediate temperature field.
    	Meanwhile, the average temperature for the current state,
    	represented as $\overline{T}^{c}$, is calculated.
    	The maximum and average residual, $e_{max}^{c}$ and 
    	$e_{ave}^{c}$ respectively, are also updated accordingly.
    	\item The PDE solving stops advancing when the 
    	current maximum residual $e_{max}^{c}$ and average residual 
    	$e_{ave}^{c}$ are both smaller than their respective values
    	obtained at the last iteration, satisfying the conditions 
    	$e_{max}^{c} \textless e_{max}^{*}$ and $e_{ave}^{c} 
    	\textless e_{ave}^{*}$.
    \end{itemize}
	\item \textbf{Iterations and termination.}
	After updating the new $ e_{max}^{*}$ 
	and $e_{ave}^{*}$, the optimization process repeats from 
	Steps 1 to 3 until the maximum residual reaches the specified 
	threshold, and the variations of temperature and thermal 
	conductivity become smaller than the thresholds. 
	Once these criteria are met, the optimization process 
	is considered complete, and the resulting distribution of $k$
	is deemed optimal.
	
\end{enumerate}

Note that the present method divides the optimization process 
into small, easily manageable steps, 
which simplifies and expedites the optimization process.
The magnitude of target strength $\beta$ is chosen as a small 
fraction of the average initial temperature,
within the range of $0.5 \sim 1$ in the present study.
In addition, the $\beta$ is adjusted dynamically,
increasing by a factor of 1.05 when the 
average temperature is lower than in the previous optimize iteration 
and decreasing with a decayed factor of 0.8 
when the temperature exceeds that of the previous iteration.
Actually, decreasing $\beta$ is quite important 
for effective convergence in the late stages of the optimization.
Furthermore, since the convexity of these optimization problems 
are not able to be established, the current method, similar to other 
general optimization approaches, does not guarantee the global optimal solution.

\subsection{Evolution of design variable}\label{sectiondesignvariableevolution}
In Step 3, the residual $e_{i}^{*}$ for particle $i$ in the PDE
is calculated as 
\begin{equation}
	e_i^* = \sum_{j}\left(k_{i}+k_{j}\right)
	\dfrac{ T_{ij}}{r_{ij}}\nabla_{i} W_{ij}V_j+\dot{Q}_{i}.
	\label{calculateoriginalresidual}
\end{equation}
Once the target is imposed on this particle, the PDE residual 
deviates from its original value and will be recovered by 
modifying the design variable $k$ on particle $i$ and 
its neighboring particles $j$.
This process can be represented by the  
pseudo-time evolution of following equation
\begin{equation}
	\frac{\text{d}k_{i}}{\text{d}\tau}=\sum_{j} 
	\left(T^{c}_{i}-T_j\right) \left(k_{i}^{m+1}+k_{j}^{m+1}\right)
	\dfrac{1}{r_{ij}}\nabla_{i} W_{ij}V_j +\dot{Q}_i + e_i^*.
\end{equation}
Here, $k_{i}^{m+1}+k_{j}^{m+1} = k_{i}^{m} + \text{d}k_{i} + k_{j}^{m} +
\text{d}k_{j}$, where $m$ is the previous time step 
and $\text{d}k_{i}$ and $\text{d}k_{j}$ 
represent increments after the new time step.
The implicit splitting operator introduced in 
Section \ref{implicit} is utilized.
Similar to the Eq. \eqref{learningrate}, 
a linear system is formed with respect to $\left(\text{d}k_1, \text{d}k_2,
\cdots, \text{d}k_{N-1}, \text{d}k_N\right)^T$.
Note that, the pseudo-time derivative on the LHS is 
essential for the stable evolution of $k$. 
If this term is omitted, the diagonal entries of the matrix 
for the linear system become
\begin{equation}
	\displaystyle\left(\eta_{1}\sum_{j}^{n}B_{1j}, 
	\eta_{2}\sum_{j}^{n}B_{2j}, \cdots, \eta_{N}\sum_{j}^{n}B_{Nj}\right)^T.
\end{equation}
It is observed that the magnitudes of diagonal entries in the 
matrix can be significantly smaller than those of non-diagonal entries,
potentially leading to numerical instability \cite{bagnara1995unified}.
On the contrast, when the pseudo-time derivative term is included,
the linear system transforms into
\begin{equation}
	\displaystyle\left(\eta_{1}\sum_{j}^{n}\left(B_{1j}-1\right), 
	\eta_{2}\sum_{j}^{n}\left(B_{2j}-1\right), \cdots, 
	\eta_{N}\sum_{j}^{n}\left(B_{Nj}-1\right)\right)^T,
\end{equation}
whose diagonal entries becomes dominant,
and therefore stabilize the evolution of the design variables.
In addition, since $k$ is a material property and should be 
non-negative, it is clipped at a lower bound of $0.0001$ 
during each iteration.

\subsection{Numerical regularization}\label{sectionregularization}
After the evolution of the design variable, it is necessary 
to apply numerical regularization, serving two essential purposes.
One is that, as previously mentioned in the Ref. \cite{zhao2022optimal}, 
the regularization plays a critical role in maintaining numerical 
stability and obtaining a smooth solution.
Secondly, it helps prevent over-fitting and avoids finding trivial local optima only.
In this study, we introduce a diffusion analogy approach for 
regularizing the distribution of the design variable, i.e. 
the thermal conductivity $k$ is treated as the variable 
again in the pseudo-time SPH discretized diffusion equation, given as
\begin{equation}
	\dfrac{\text{d}k_{i}}{\text{d}\tau} = 2 \sum_{j}\mu 
	\dfrac{k_{ij}}{r_{ij}}\nabla_{i} W_{ij}V_j,
	\label{regularizationformulation}
\end{equation}
where $k_{ij} = k_{i} - k_{j}$ 
and $\mu$ is the artificial diffusion coefficient used to control 
the rate of regularization.
We choose $\mu$ to be general in the range $1 \sim 2$ 
according to the target strength.
The coefficient also undergoes a similar dynamical adjustment strategy 
as the target strength because smaller $\beta$ 
usually require less regularization to achieve a smooth field.
Note that the pseudo-time derivative term is also used in
Eq. \eqref{regularizationformulation} 
to ensure that the diagonal is dominant.
%
%%%%%%%%%%%%%%%%%%%%%%%%%%%%%%%%%%%%%%%%%%%%%%%%%%%%%%%%%%%%%
%
% Section
%
%%%%%%%%%%%%%%%%%%%%%%%%%%%%%%%%%%%%%%%%%%%%%%%%%%%%%%%%%%%%%
\section{Results and discussion}\label{results}
%%%%%%%%%%%%%%%%%%%%%%%%%%%%%%%%%%%%%%%%%%%%%%%%%%%%%%%%%%%%%
In this section, we present a set of example problems to 
validate the effectiveness of the proposed target-driven 
PDE-constrained optimization method (referred to as the 
TD) for optimizing the thermal conductivity distribution 
within a heat conduction domain.

\subsection{The 2/10 heat sinks with uniform internal heat source}
\begin{figure}[htbp]
	\centering
	\makebox[\textwidth][c]{\subfigure[]{
			\includegraphics[width=.45\textwidth]{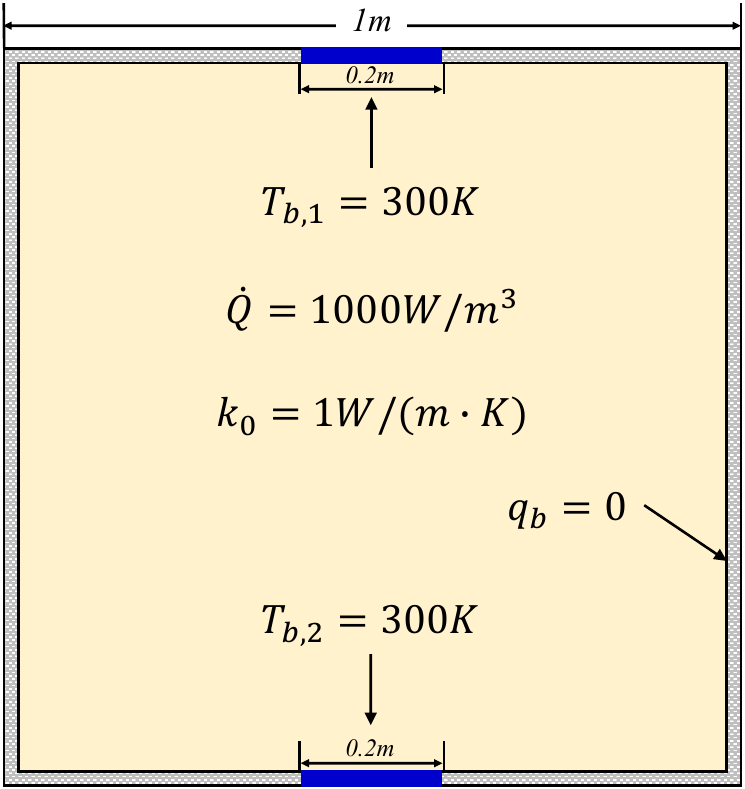}
			\label{problem1_set_up}
		}
		\quad
		\subfigure[]{
			\includegraphics[width=.45\textwidth]{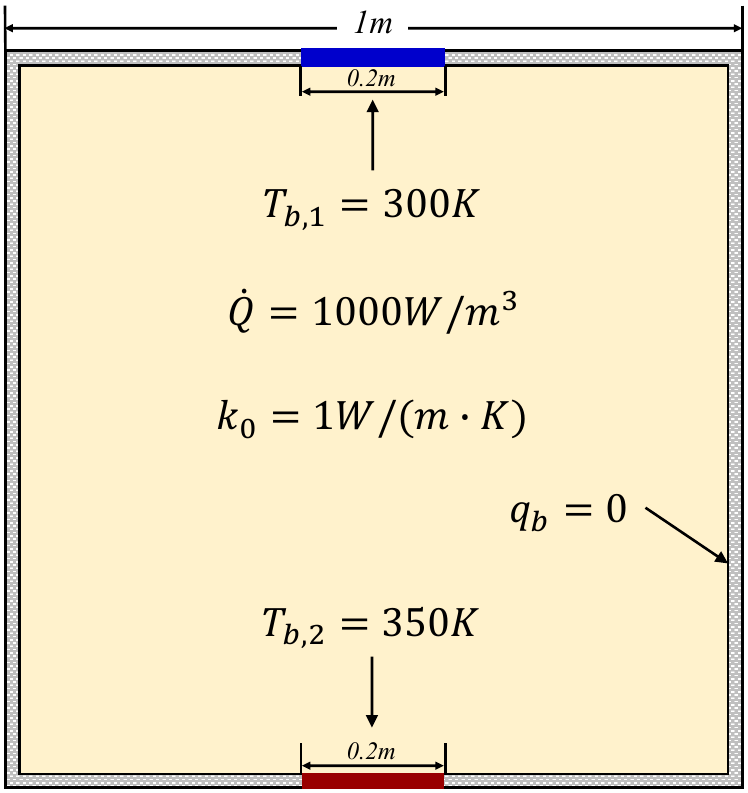}
			\label{problem2_set_up}
		}
		\quad}
	\makebox[\textwidth][c]{\subfigure[]{
			\includegraphics[width=.45\textwidth]{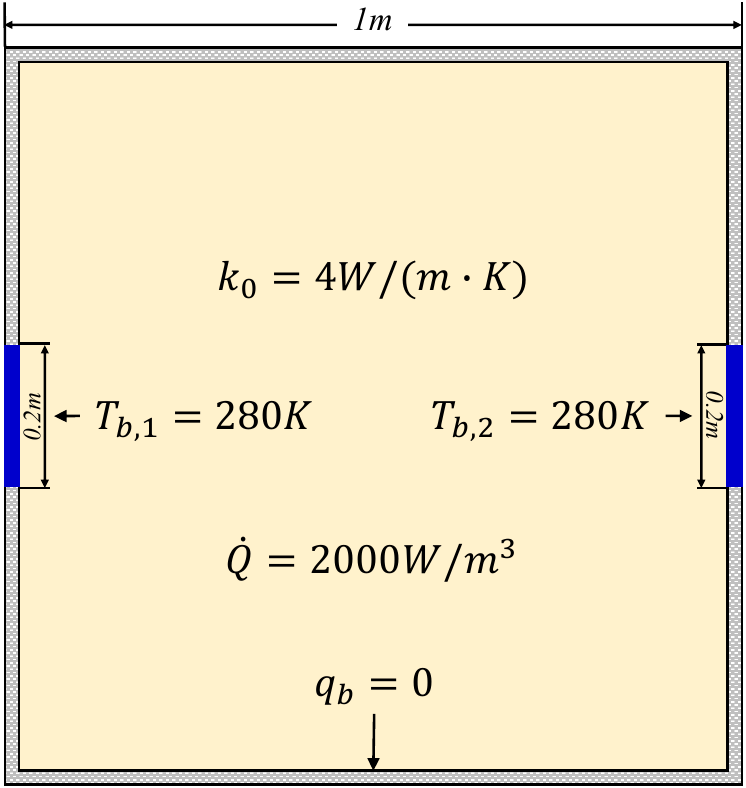}
			\label{problem3_set_up}
		}
		\quad
		\subfigure[]{
			\includegraphics[width=.45\textwidth]{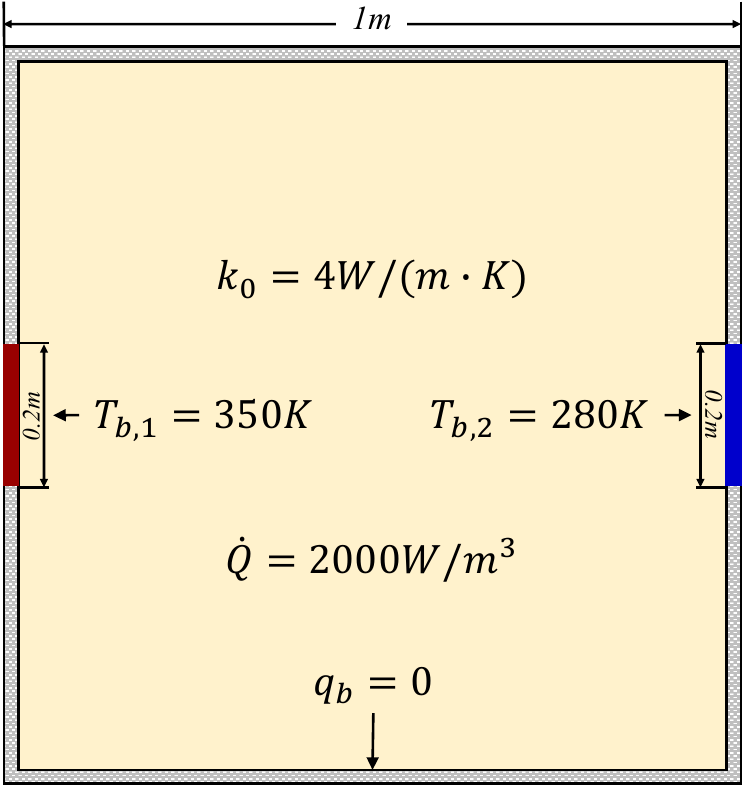}
			\label{problem4_set_up}
		}
		\quad}
	\caption{Illustration of the problem setups.
		(a) Problem 1: Identical heat sinks; 
		(b) Problem 2: Non-identical heat sinks;
		(c) Problem 3: Identical heat sinks with higher $k_{0}$;
		(d) Problem 4: Non-identical heat sinks with higher $k_{0}$.}
	\label{section1_setup}
\end{figure}

The first set of problems involves a square thermal domain measuring 
1m on each side, with a uniform internal heat source.   
Two heat sinks, each covering 20\% of the side length, 
are symmetrically positioned at the center of opposite boundaries,
maintaining the constant temperature.
The remaining boundaries are set as adiabatic.
Various scenarios are considered, including different heat source intensities, 
initial thermal conductivity values, 
and cases with either identical or non-identical sink temperatures.
As shown in Fig. \ref{section1_setup},
the detailed setups for these problems are illustrated as follows:
\begin{itemize}
	\item Problems 1 and 2 feature a uniform heat source with an intensity 
	of $\rm 1000W/m^{3}$, while Problems 3 and 4 have a uniform heat source 
	with an intensity of $\rm 2000W/m^{3}$.
	\item The initial thermal conductivity, denoted as $k_0$, is 
	1W/(m·K) for Problems 1 and 2, while it is 4W/(m·K) for Problems 3 and 4.
	\item Problems 1 and 3 have identical heat sink temperatures,
	set at 300K and 280K, respectively.
	Problems 2 and 4 involve non-identical heat sink temperatures,
	with one of the heat sinks reaching a higher temperature of 350K.
\end{itemize}
The thermal domain is discretized with 100 particles on each side, 
resulting in a total of 10,000 particles.
Additionally, four layers of dummy particles are used to 
enforce boundary conditions. 
Note that the problems discussed in the following sections share
the same configuration for the SPH implementation.
Reference solutions for Problems 1 and 2 were obtained using 
automatic differentiation (AD) and the temperature gradient 
homogenization (TGH) method, and can be found in 
Ref. \cite{song2021optimization}.
Reference solutions for Problems 3 and 4 were achieved through 
adjoint analysis (AA) and the TGH method, and are available in
Ref. \cite{zhao2022optimal}.

\begin{figure}[htbp]
	\centering
	\makebox[\textwidth][c]{\subfigure[]{
			\includegraphics[width=.45\textwidth]{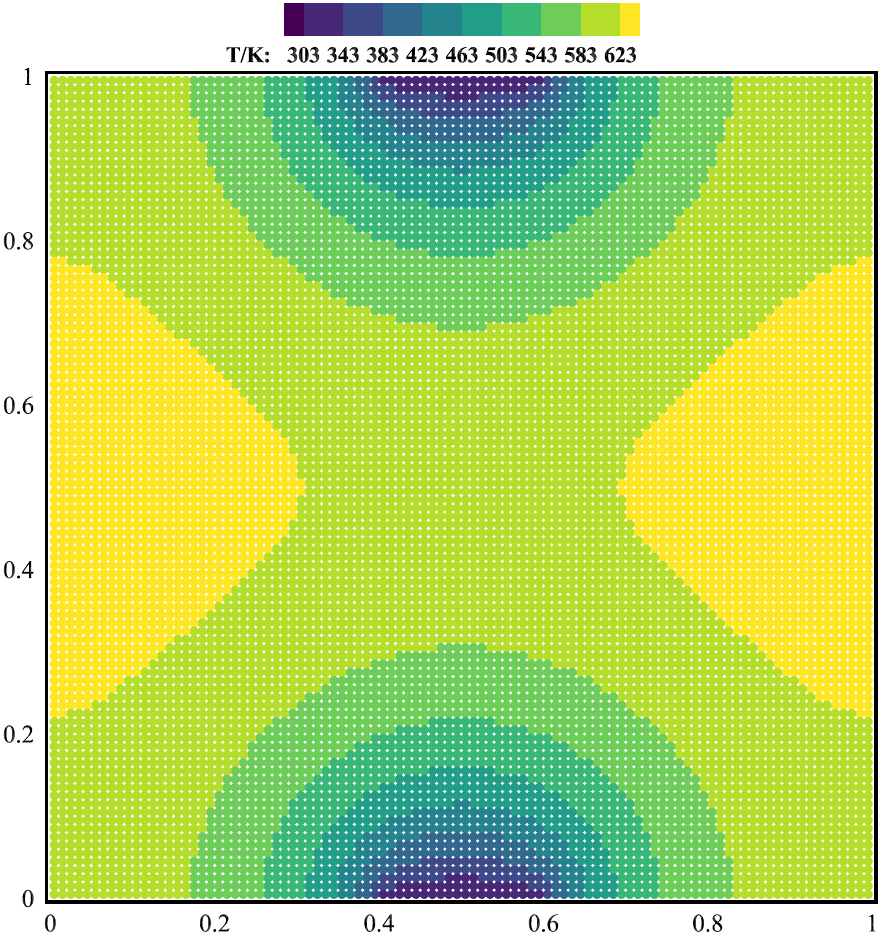}
			\label{problem1_non_tem}
		}
		\quad
		\subfigure[]{
			\includegraphics[width=.45\textwidth]{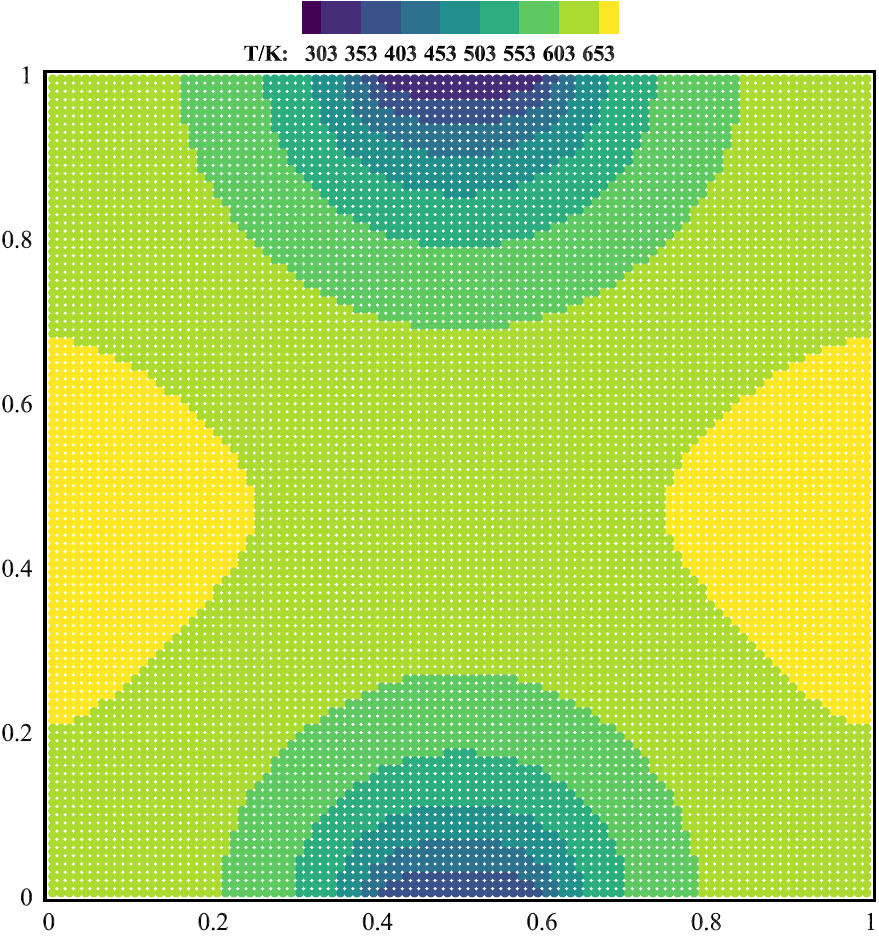}
			\label{problem2_non_tem}
		}
		\quad}
	\makebox[\textwidth][c]{\subfigure[]{
			\includegraphics[width=.45\textwidth]{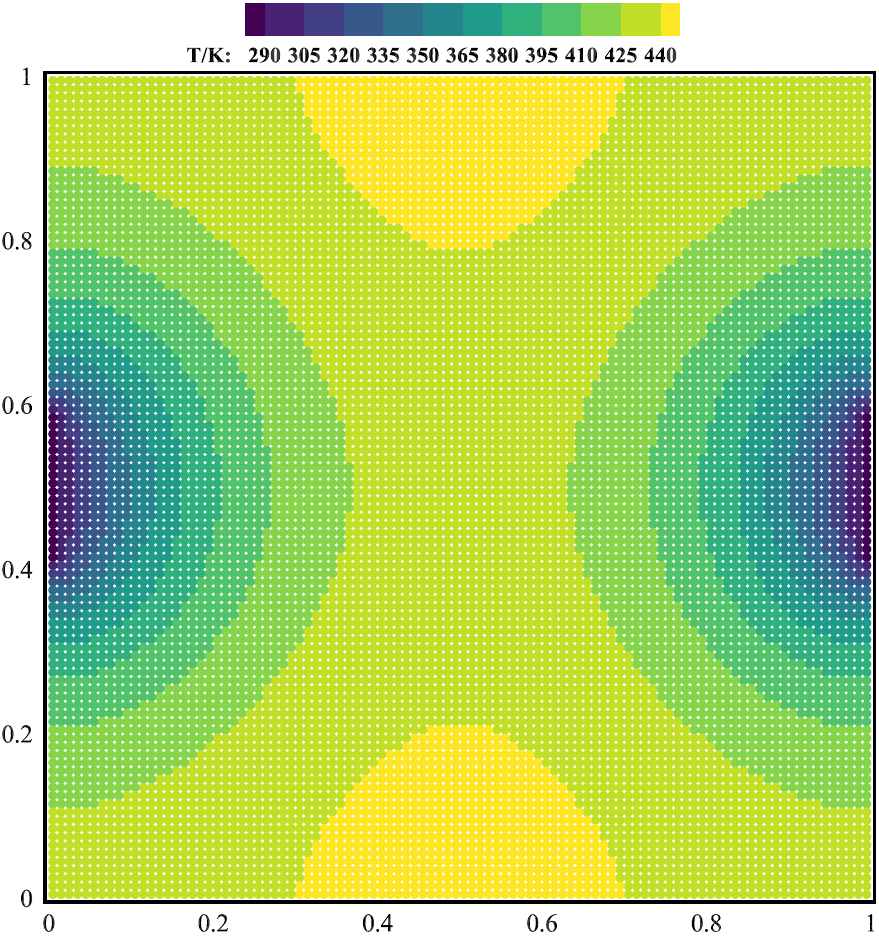}
			\label{problem3_non_tem}
		}
		\quad
		\subfigure[]{
			\includegraphics[width=.45\textwidth]{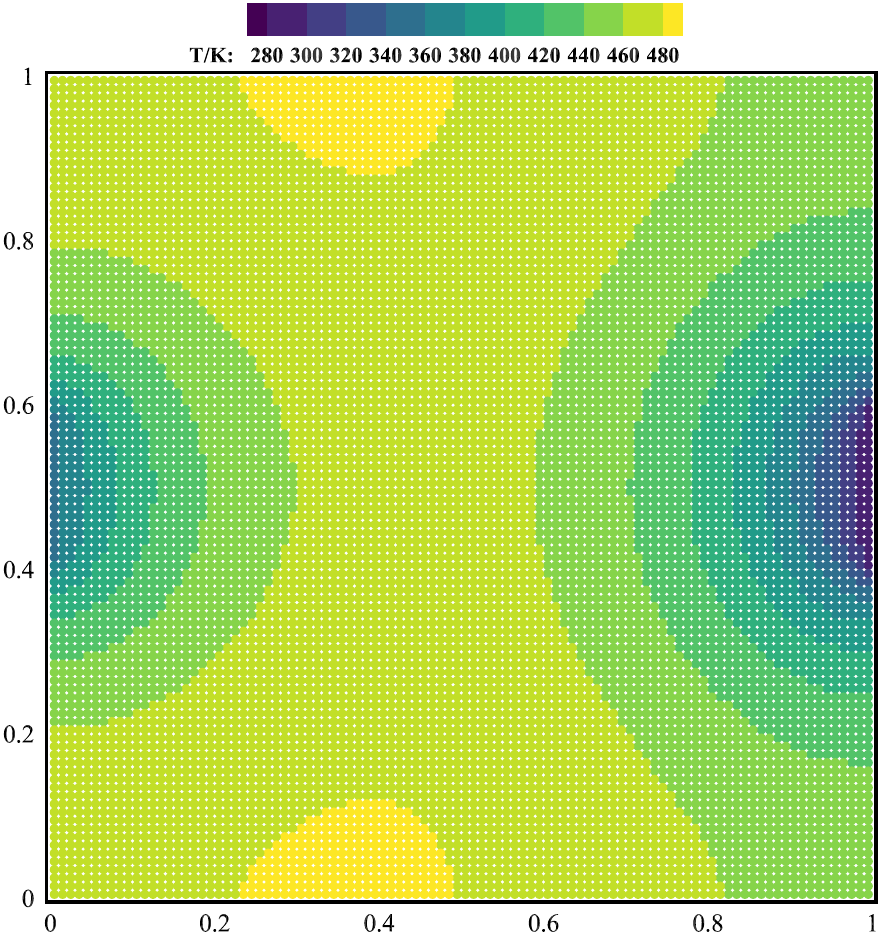}
			\label{problem4_non_tem}
		}
		\quad}
	\caption{Temperature distributions with uniform thermal conductivity 
		for different scenarios.
		(a) Problem 1; (b) Problem 2;
		(c) Problem 3; (d) Problem 4.}
	\label{section1_non_tem}
\end{figure}

The steady temperature distributions with uniform thermal conductivity 
are presented in Fig. \ref{section1_non_tem}.
All the obtained distributions are in good agreement with the reference 
results (see Figs. 4 and 10 of Ref. \cite{song2021optimization}).
In these cases, heat generated by the internal source is conducted 
throughout the domain and subsequently dissipated at the heat sinks. 
A more pronounced temperature gradient is observed near the heat sinks,
indicating a concentration of heat flux and resulting in higher 
temperature within the domain.
The average temperatures obtained are slightly higher than that in 
the reference, which may be attributed to variations in 
resolution and discretization employed.

\begin{figure}[htbp]
	\centering
	\makebox[\textwidth][c]{\subfigure[]{
			\includegraphics[width=.45\textwidth]{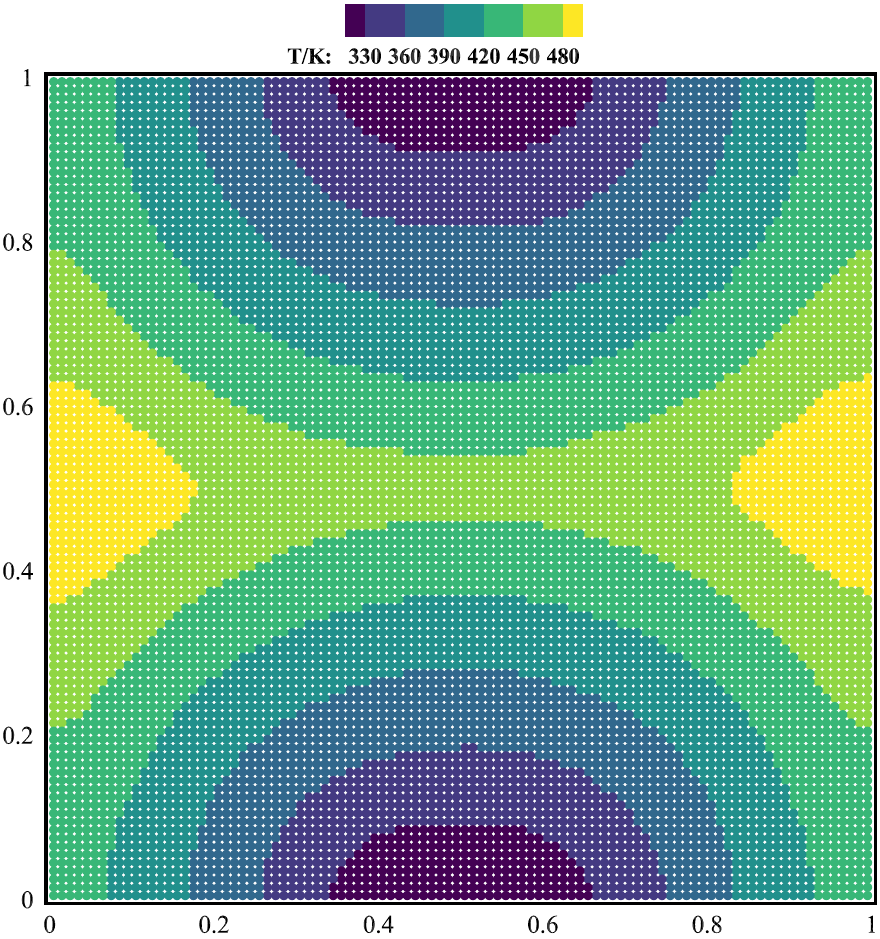}
			\label{problem1_op_tem_td}
		}
		\quad
		\subfigure[]{
			\includegraphics[width=.45\textwidth]{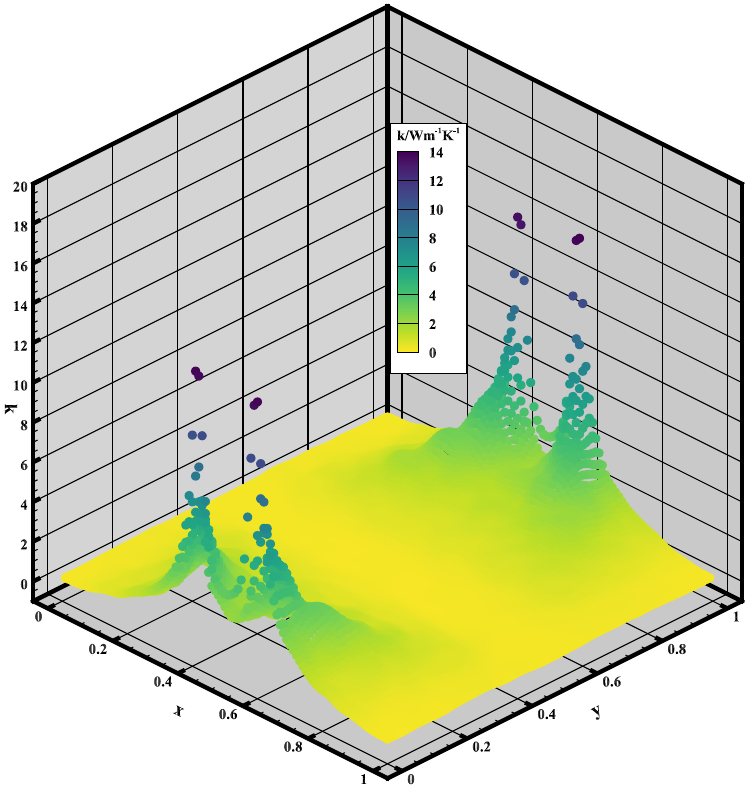}
			\label{problem1_op_k_td}
		}
		\quad}
	\caption{Present optimized results of the Problem 1. 
		(a) Temperature; (b) Thermal conductivity.}
	\label{problem1_op_tem}
\end{figure}

\begin{figure}[htbp]
	\centering
	\makebox[\textwidth][c]{\subfigure[]{
			\includegraphics[width=.45\textwidth]{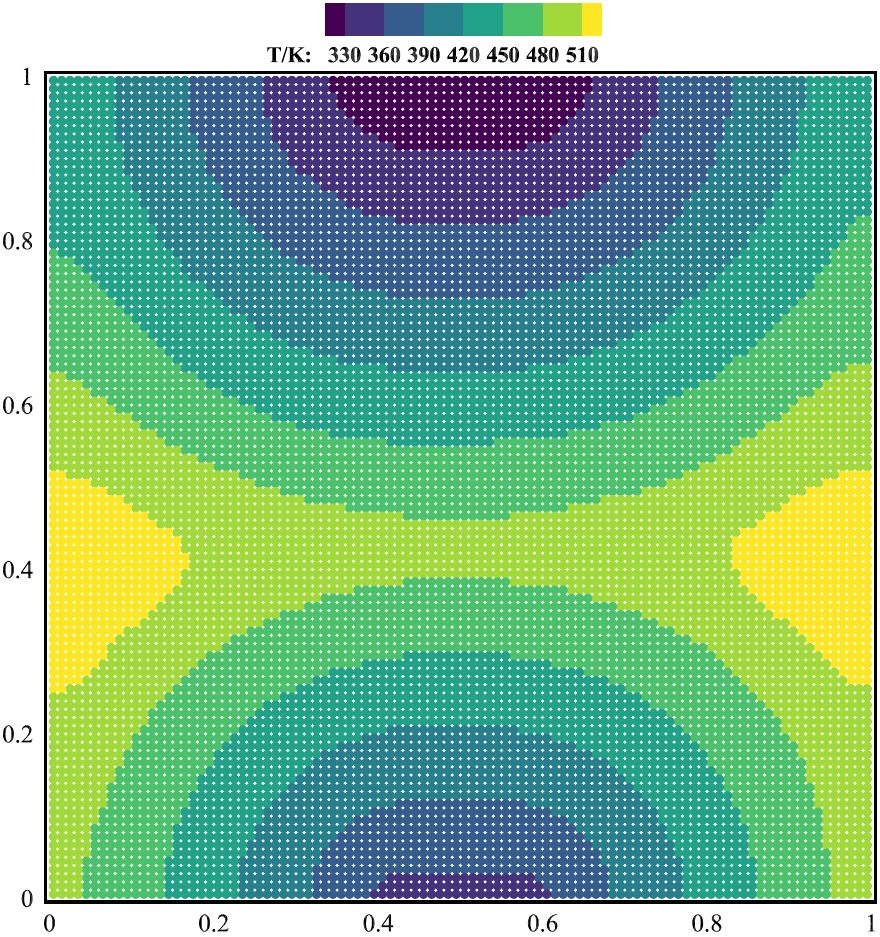}
			\label{problem2_op_tem_td}
		}
		\quad
		\subfigure[]{
			\includegraphics[width=.45\textwidth]{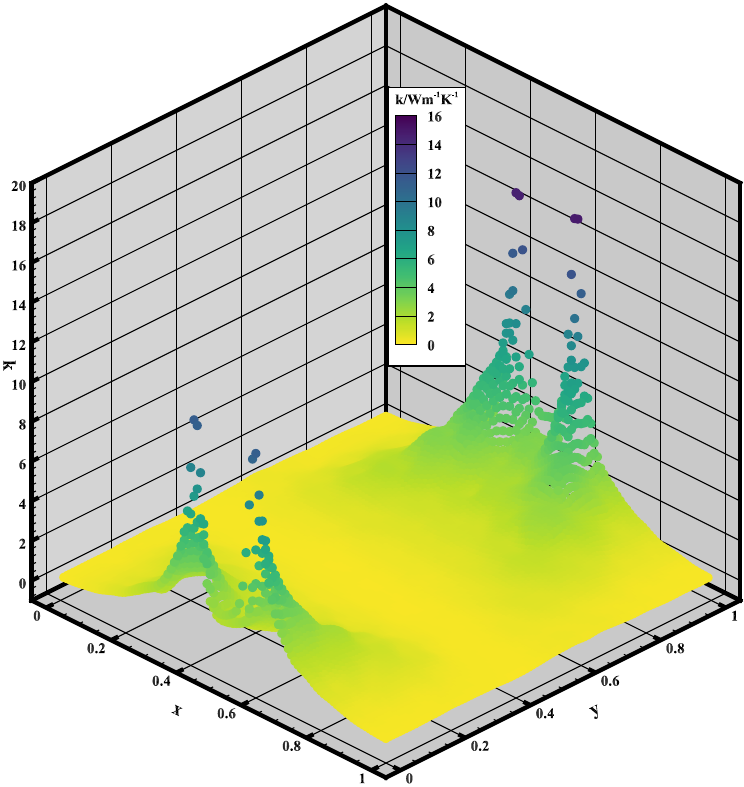}
			\label{problem2_op_k_td}
		}
		\quad}
	\caption{Present optimized results of the Problem 2.
		(a) Temperature; (b) Thermal conductivity.}
	\label{problem2_op_tem}
\end{figure}

\begin{figure}[htbp]
	\centering
	\makebox[\textwidth][c]{\subfigure[]{
			\includegraphics[width=.45\textwidth]{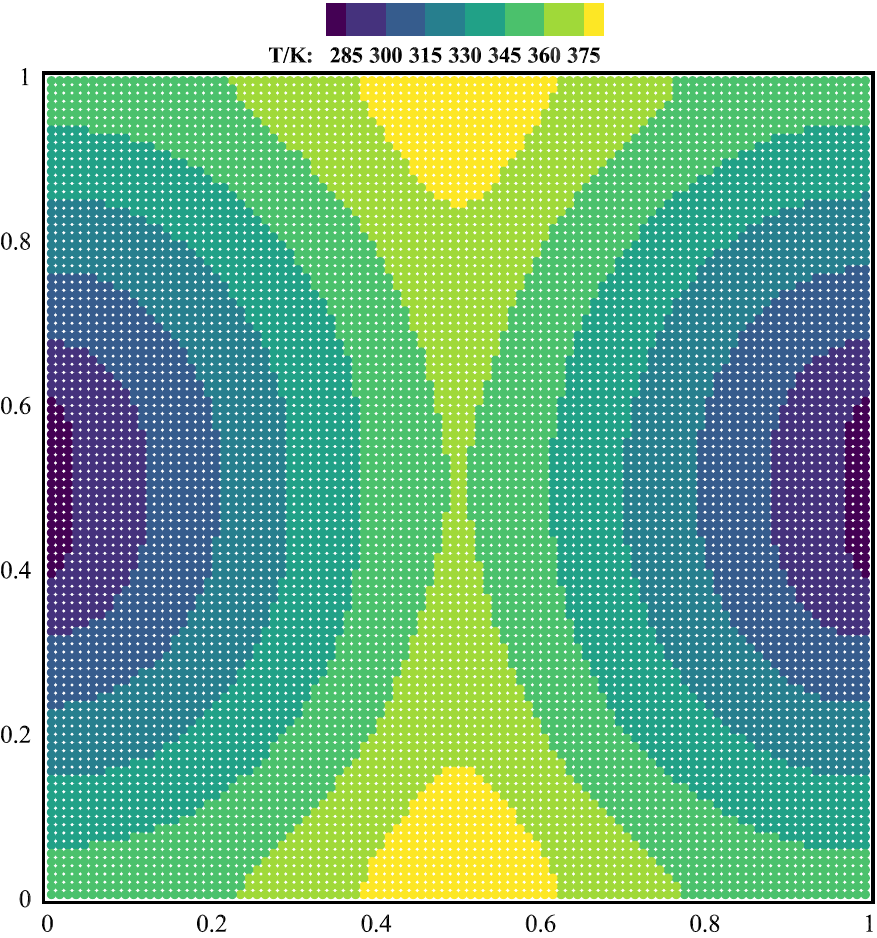}
			\label{problem3_op_tem_td}
		}
		\quad
		\subfigure[]{
			\includegraphics[width=.45\textwidth]{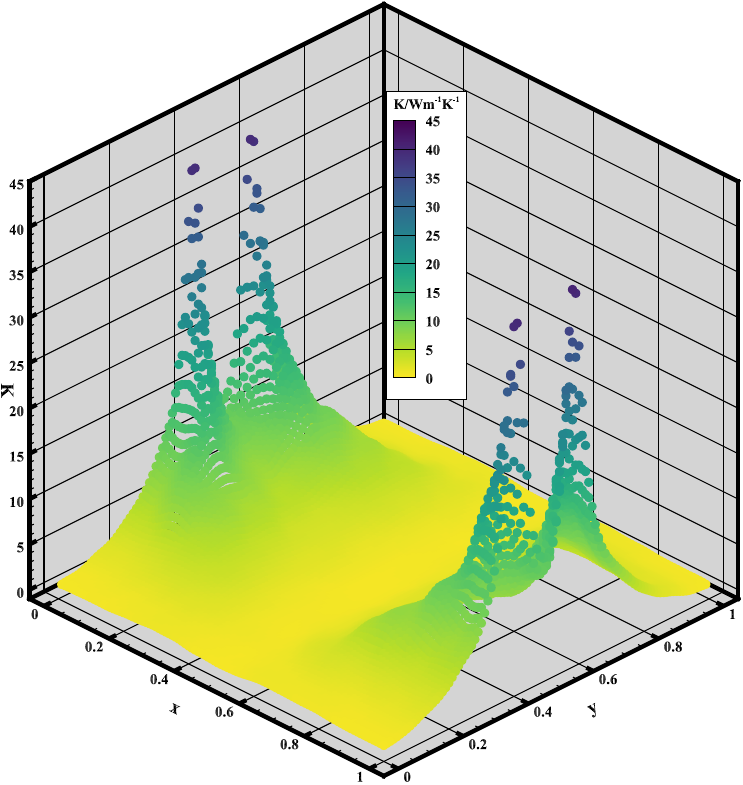}
			\label{problem3_op_k_td}
		}
		\quad}
	\caption{Present optimized results of the Problem 3.
		(a) Temperature; (b) Thermal conductivity.}
	\label{problem3_op_tem}
\end{figure}

\begin{figure}[htbp]
	\centering
	\makebox[\textwidth][c]{\subfigure[]{
			\includegraphics[width=.45\textwidth]{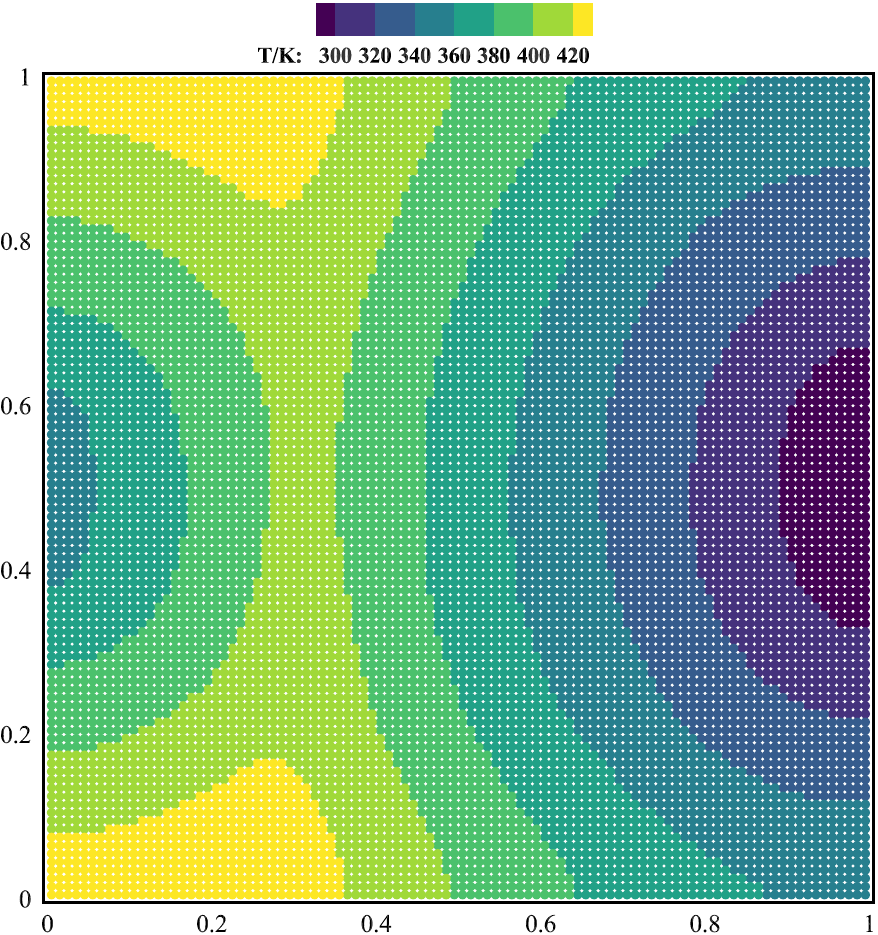}
			\label{problem4_op_tem_td}
		}
		\quad
		\subfigure[]{
			\includegraphics[width=.45\textwidth]{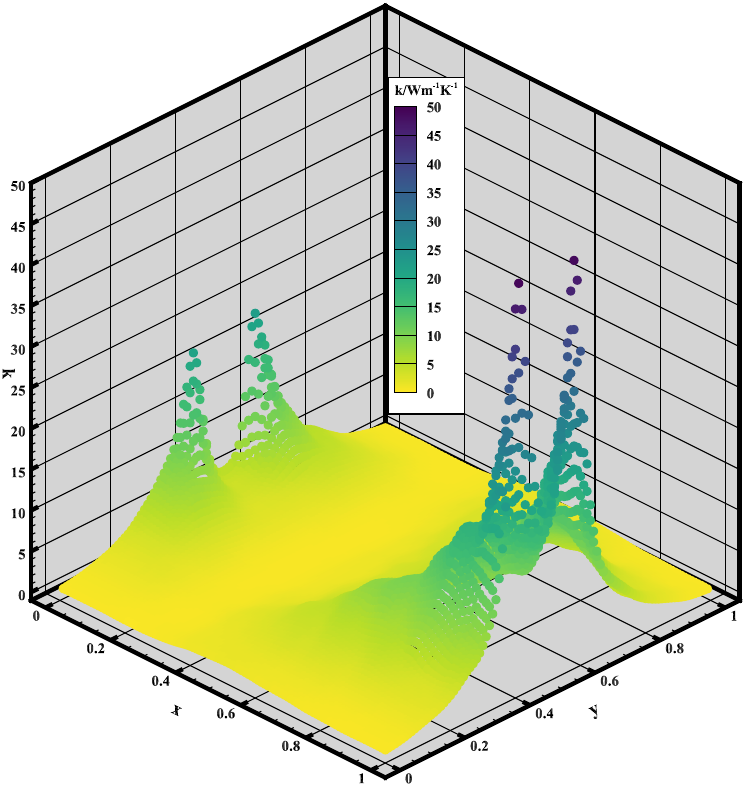}
			\label{problem4_op_k_td}
		}
		\quad}
	\caption{Present optimized results of the Problem 4.
		(a) Temperature; (b) Thermal conductivity.}
	\label{problem4_op_tem}
\end{figure}

For these problems, the optimization results obtained by
the present method are displayed in Figs. \ref{problem1_op_tem} to \ref{problem4_op_tem},
and the comparisons with previous work are summarized in Tables
\ref{problem1_3_comparison_temperature} and 
\ref{problem2_4_comparison_temperature}.
Problems 1 and 3, featuring identical heat sinks, result in a symmetrical 
pattern in the optimized results that align with the reference results.
In cases where the heat sinks are identical, the problem becomes 
self-adjoint, making the TGH method equivalent to the AA method 
when the target is to minimize the global temperature \cite{alexandersen2021revisiting}.
Therefore, all methods adopt the direct target and exhibit
similar optimal performance, although the reference result for 
the Problem 1 suggests slightly superior outcome using the AD method.
Notably, the present method tends to yield a higher reduction 
in temperature for these two problems.
The optimization process homogenized the temperature gradient 
to balance heat flux through the domain.
Optimized temperature distributions in Figs. \ref{problem1_op_tem_td} 
and \ref{problem3_op_tem_td} reveal an overall even temperature 
gradient, matching with the reference results obtained by other methods
(see Fig. 5(b) of Ref. \cite{song2021optimization} 
 and Fig. 6(a) of Ref. \cite{zhao2022optimal}). 
Furthermore, the optimization process results in a significant 
increase in thermal conductivity near the heat sinks. 
The distribution of optimized thermal conductivity in 
Figs. \ref{problem1_op_k_td} and \ref{problem3_op_k_td} 
exhibits four distinct peaks prominently located at the 
edges of the heat sinks, closely matching the reference 
(see Fig. 5(a) of Ref. \cite{song2021optimization}
 and Fig. 6(b) of Ref. \cite{zhao2022optimal}).
These peaks effectively interact with other boundaries, 
enhancing the efficient dissipation of generated heat.
It's worth noting, however, that the peak thermal conductivity obtained 
by the present method approximates 14 W/(m·K) and 40 W/(m·K) for 
Problems 1 and 3, respectively, which are considerable less than the 
values in the reference (beyond 20 W/(m·K) and 50 W/(m·K)).
\begin{table}[htbp]
	\small
	\renewcommand\arraystretch{1.25}
	\centering
	\captionsetup{font={small}}
	\caption{Summary of result comparisons for problems with identical sinks.}
	\begin{tabularx}{14cm}{@{\extracolsep{\fill}}ccccc}
		\hline
		\quad Problem &Method & Original $\overline{T}\left(K\right)$ 
		& Optimized $\overline{T}\left(K\right)$ 
		& Reduced $\left(\% \right)$ \quad\\
		\midrule
	    \quad \multirow{3}{*}{1} 
	    &TD                                 & $584.30$ & $413.88$ & $29.17$ \quad\\
		&AD \cite{song2021optimization}     & $582.04$ & $412.73$ & $29.09$ \quad\\
		&TGH \cite{song2021optimization}    & $582.04$ & $413.45$ & $28.97$ \quad\\
		\midrule
		\quad \multirow{3}{*}{3}
		&TD method                          & $417.59$ & $339.12$ & $18.79$ \quad\\
		&AA \cite{zhao2022optimal}          & $414.36$ & $336.98$ & $18.67$ \quad\\
		&TGH \cite{zhao2022optimal}         & $414.36$ & $336.99$ & $18.67$ \quad\\
		\bottomrule
&	\end{tabularx}
	\label{problem1_3_comparison_temperature}
\end{table}
\begin{table}[htbp]
	\small
	\renewcommand\arraystretch{1.25}
	\centering
	\captionsetup{font={small}}
	\caption{Summary of result comparisons for problems with non-identical sinks.}
	\begin{tabularx}{14cm}{@{\extracolsep{\fill}}ccccc}
		\hline
		\quad Problem &Method & Original $\overline{T}\left(K\right)$ 
		& Optimized $\overline{T}\left(K\right)$ 
		& Reduced $\left(\% \right)$ \quad\\
		\midrule
		\quad \multirow{3}{*}{2}
		&TD                                 & $609.30$ & $438.79$ & $27.98$ \quad\\
		&AD  \cite{song2021optimization}    & $607.04$ & $436.96$ & $28.02$ \quad\\
		&TGH \cite{song2021optimization}    & $607.04$ & $438.20$ & $27.81$ \quad\\
		\midrule
		\quad \multirow{3}{*}{4}
		&TD                                 & $452.73$ & $371.66$ & $17.91$ \quad\\
		&AA \cite{zhao2022optimal}          & $449.36$ & $368.61$ & $17.97$ \quad\\
		&TGH \cite{zhao2022optimal}         & $449.36$ & $371.55$ & $17.32$ \quad\\
		\bottomrule
		&	\end{tabularx}
	\label{problem2_4_comparison_temperature}
\end{table}
The differences in heat sink configurations in Problems 2 and 
4 highlight that the AA and TGH methods are no longer equivalent.
The TGH method transitions into an indirect target approach, 
while the direct target methods, TD, AD, and AA, consistently 
outperform the TGH.
The present method exhibits a slightly lower temperature reduction 
ratio compared to the other two direct target methods. 
Interestingly, when we decrease the artificial regularization 
coefficient ($\mu$), 
it tends to yield a lower temperature, albeit within certain limits.
However, this adjustment results in significantly higher peak 
values of the thermal conductivity. 
Therefore, we have chosen the current coefficient value 
for a balance between optimization performance and 
the peak value of thermal conductivity.
This choice provides greater flexibility in selecting high thermally conductive 
but electrically isolated materials for cooling electronic devices without 
sacrificing optimal performance.
The optimized temperature distributions in Figs. \ref{problem2_op_tem_td} 
and \ref{problem4_op_tem_td} exhibit a generally even gradient, matching
with the reference (see Fig. 11(b) of Ref. \cite{song2021optimization} 
and Fig. 7(a) of Ref. \cite{zhao2022optimal}).
Similarly, the optimized thermal conductivity distributions in 
Figs. \ref{problem2_op_k_td} and \ref{problem4_op_k_td} still feature 
four peaks around the sinks.
However, the heights of opposite peaks are no longer equal 
due to temperature discrepancies, 
with larger heights observed near the colder sinks.
These features closely align with the reference results
(see Fig. 11(a) of Ref. \cite{song2021optimization} 
and Fig. 7(b) of Ref. \cite{zhao2022optimal}), and notably, 
the present method is always characterized with lower peaks.

All simulations and optimizations were performed on 
a computer equipped with 2 Intel(R) Xeon(R) CPU 
E5-2680 v4 processors.
For Problems 1 to 4, it takes approximately 100 seconds 
to obtain a steady-state temperature field, and 
detailed information regarding the optimization duration
is presented in Table \ref{time-consuming}.
\begin{table}[htbp]
	\small
	\renewcommand\arraystretch{1.25}
	\centering
	\captionsetup{font={small}}
	\caption{Summary of the optimization process for Problems 1 to 4.}
	\begin{tabularx}{14cm}{@{\extracolsep{\fill}}cccccc}
		\hline
		\quad \multirow{2}{*}{Problem} &\multirow{2}{*}{Steady time(s)} 
		&\multicolumn{3}{c}{Optimized iteration} &\multirow{2}{*}{Ratio} \quad\\ \cline{3-5} 
		& &Loop &Step &Time(s)\\
		\midrule
		\quad 1 & $102.8$ & $108$ & $160549$ & $264.3$ & $2.6$ \quad\\ 
		\quad 2 & $100.6$ & $94 $ & $160541$ & $265.4$ & $2.6$ \quad\\
		\quad 3 & $110.2$ & $106$ & $210906$ & $341.1$ & $3.1$ \quad\\
		\quad 4 & $122.1$ & $104$ & $169990$ & $279.5$ & $2.3$ \quad\\
		\bottomrule
	\end{tabularx}
	\label{time-consuming}
\end{table}
Note that,
although the actual optimization time for each run 
may vary and be depending on the selected parameters,
the shown results suggest that
the present method is quite efficient,
as it only takes a few times the computation cost 
of obtaining a steady-state solution to achieve the optimized result.  
\subsection{The 2/10 sinks with Gaussian distributed heat source}
\begin{figure}[htbp]
	\centering
	\subfigure[]{\includegraphics[width=.45\textwidth]{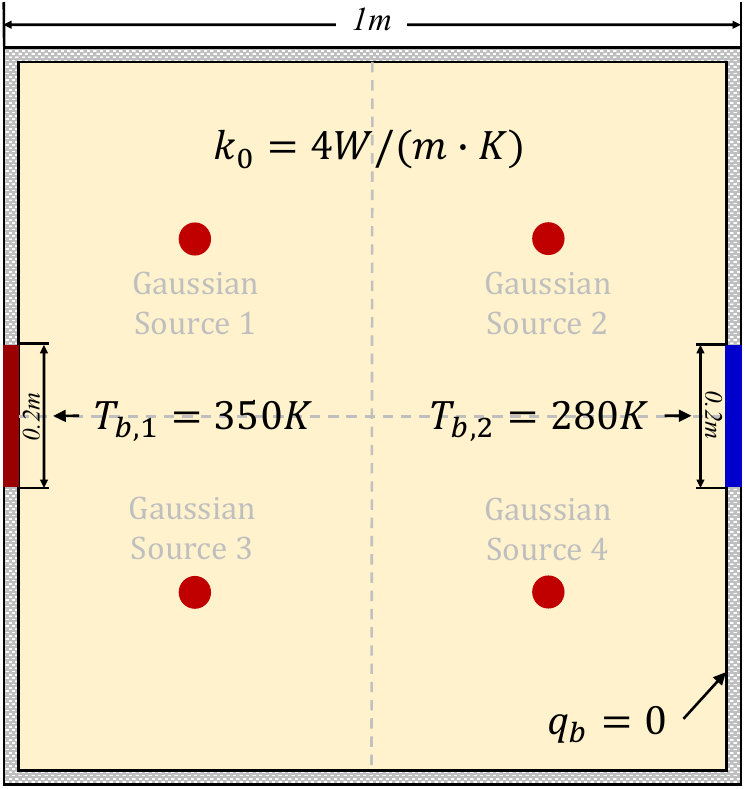}
		\label{problem5_sketch}
	}
	\makebox[\textwidth][c]{\subfigure[]{
			\includegraphics[width=.45\textwidth]{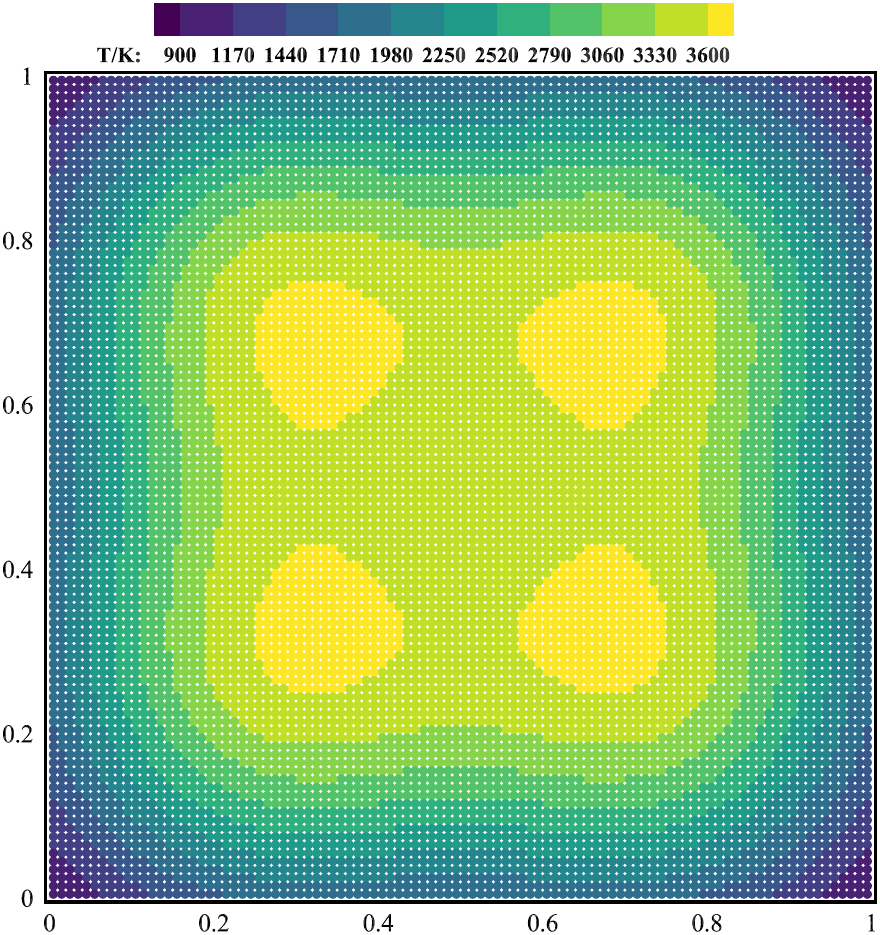}
			\label{problem5_heat_source}
		}
		\quad
		\subfigure[]{
			\includegraphics[width=.45\textwidth]{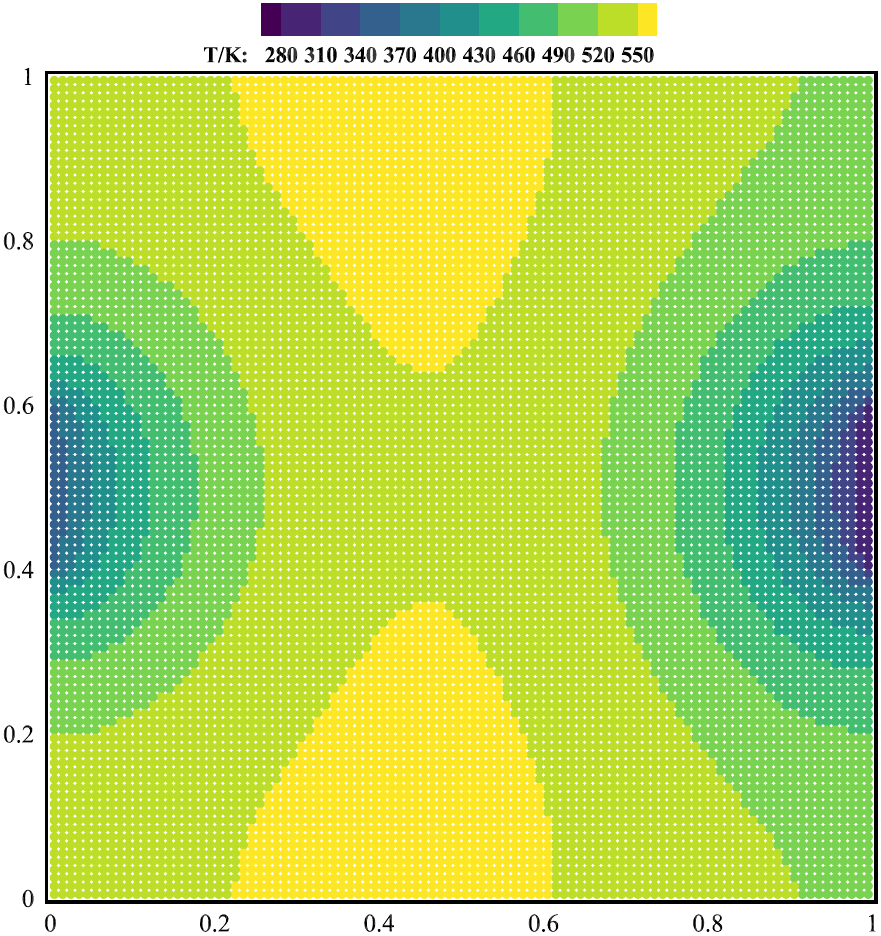}
			\label{problem5_non_tem}
		}
		\quad}
	\caption{Problem 5. 
		(a) Schematic depiction of the problem setup;
		(b) Overall heat source intensity;
		(c) Temperature distribution under uniform thermal conductivity.}
	\label{problem5_set_up}
\end{figure}

Problem 5 closely resembles Problem 4; however, it introduces a 
non-uniform distribution of the heat source.
Four Gaussian heat sources are symmetrically placed within the 
domain at four coordinates: (0.25, 0.25), (0.25, 0.75), 
(0.75, 0.25), and (0.75, 0.75), as depicted in Fig. \ref{problem5_sketch}.
The heat sources $\dot{Q}_{i}$ are determined by
\begin{equation}
	\dot{Q}_{i} = C_{i}exp\left[-10\left(\left(x-x_{i}\right)^2
	+\left(y-y_{i}\right)^2\right)\right].
\end{equation}
Here $\left(x_{i}, y_{i}\right)$ represents the center point of 
each heat source, and $C_{i}\rm=3000 W/m^{3}$ denotes the intensity.
Fig. \ref{problem5_heat_source} illustrates the cumulative heat 
source intensity across the entire domain.
Reference solutions obtained by the AA and TGH methods are also 
available in Ref. \cite{zhao2022optimal}.
The steady temperature distribution with uniform thermal conductivity 
is shown in Fig. \ref{problem5_non_tem}, and the obtained average 
temperature is 518.55K, which agrees well with the reference.

\begin{figure}[htbp]
	\centering
	\makebox[\textwidth][c]{\subfigure[]{
			\includegraphics[width=.45\textwidth]{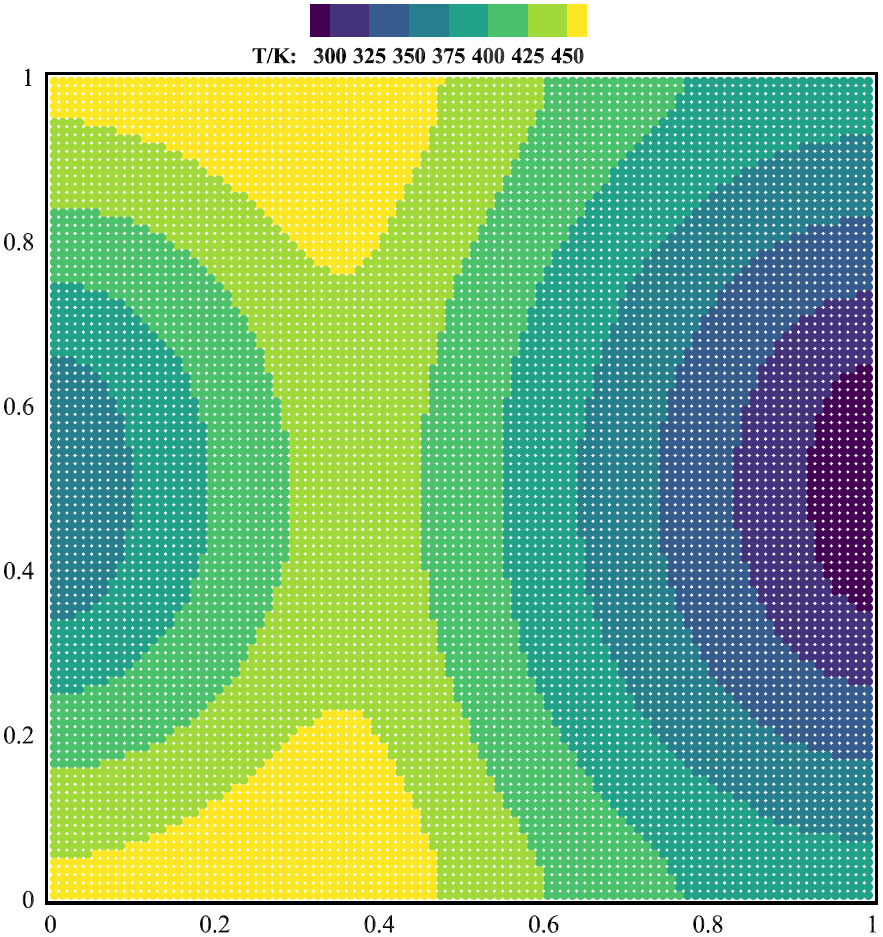}
			\label{problem5_op_tem_td}
		}
		\quad
		\subfigure[]{
			\includegraphics[width=.45\textwidth]{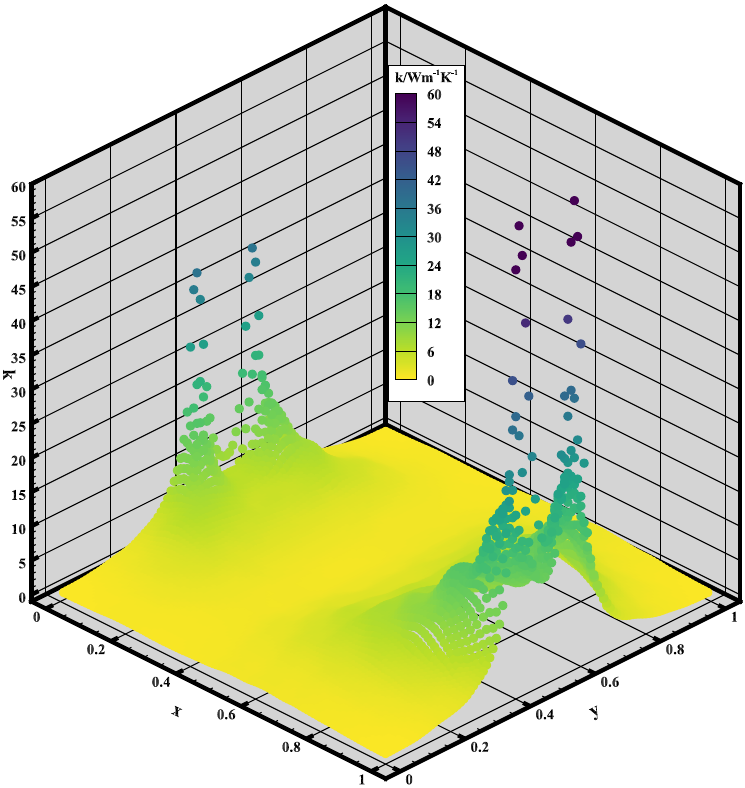}
			\label{problem5_op_k_td}
		}
		\quad}
	\caption{Present optimized results of the Problem 5.
		(a) Temperature; (b) Thermal conductivity.}
	\label{problem5_op_tem}
\end{figure}

The current optimization results are presented in Fig. \ref{problem5_op_tem},
and summarizes the result comparisons in Table \ref{problem5_comparison}.
Due to the non-identical heat sink configuration, it's evident 
that the TGH method is no longer equivalent to the AA method, 
and the reference results demonstrate that AA still outperforms 
the latter.
By employing a direct target, the present method further explores 
the optimal results and yields more temperature reduction, 
being aligned with the performance of the AA method,
as it is also based on the direct target. 
The optimized temperature contour shown in Fig. \ref{problem5_op_tem_td} 
reveals a notably lower values, particularly on 
the colder heat sink side, indicating an enhanced cooling capacity.
The temperature gradient is more evenly distributed, 
aligning with the reference results (their Fig. 10(a)).
Additionally, the optimized distribution in 
Fig. \ref{problem5_op_k_td} continues to feature the characteristic 
four peaks similar to the reference (their Fig. 10(b)).
However, the present thermal conductivity increases 
continuously around the heat sink region, 
as opposed to being concentrated at isolated spots as in the reference.
In addition, the simulation time for a steady solution is 112.4 seconds for this case, 
while the optimization time is 980.7 seconds, 
again indicating quite good efficiency.
\begin{table}[htbp]
	\small
	\renewcommand\arraystretch{1.25}
	\centering
	\captionsetup{font={small}}
	\caption{Summary of result comparisons for Problem 5.}
	\begin{tabularx}{14cm}{@{\extracolsep{\fill}}ccccc}
		\hline
		\quad Problem &Method & Original $\overline{T}\left(K\right)$ 
		& Optimized $\overline{T}\left(K\right)$ 
		& Reduced $\left(\% \right)$ \quad\\
		\midrule
		\quad \multirow{3}{*}{5}
		&TD                                 & $518.55$ & $400.33$ & $22.80$ \quad\\
		&AA \cite{zhao2022optimal}          & $517.61$ & $417.35$ & $19.37$ \quad\\
		&TGH \cite{zhao2022optimal}         & $517.61$ & $422.85$ & $18.31$ \quad\\
		\bottomrule
	\end{tabularx}
	\label{problem5_comparison}
\end{table}
\subsection{The 1/10 heat sinks with uniform internal heat source}
\begin{figure}
	\makebox[\textwidth][c]{\subfigure[]{
		\includegraphics[width=.45\textwidth]{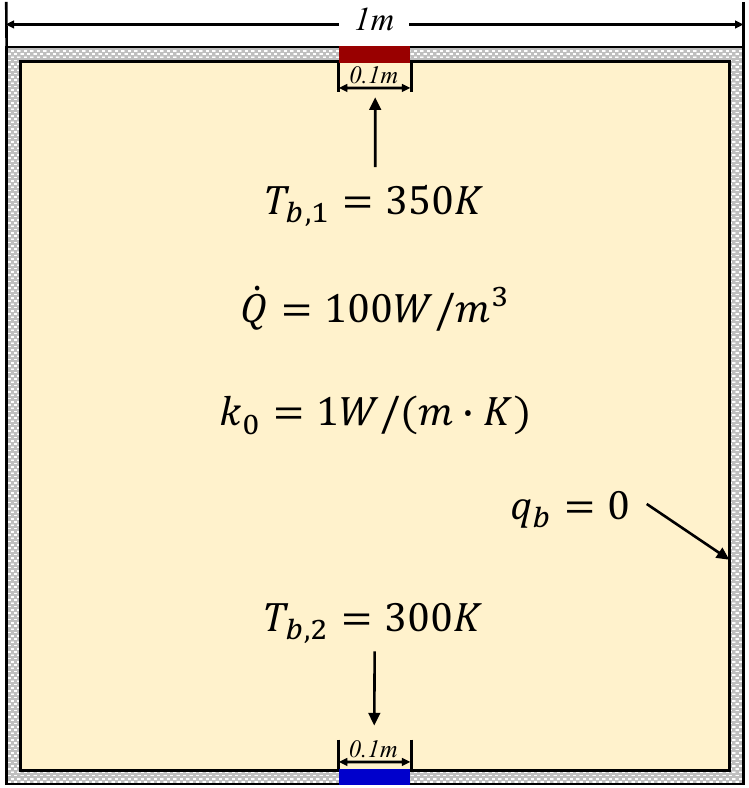}
		\label{problem6_set_up}
	}
	\quad
	\subfigure[]{
		\includegraphics[width=.45\textwidth]{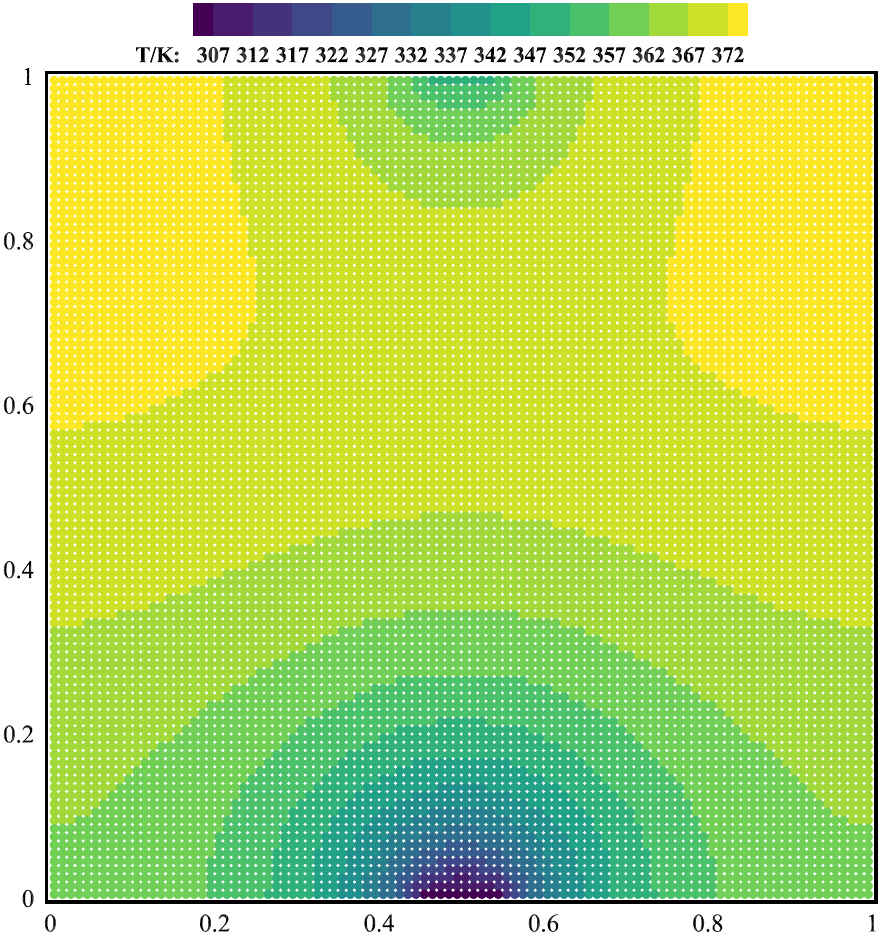}
		\label{problem6_non_tem}
	}
	\quad}
\caption{Problem 6. 
	     (a) Schematic depiction of the problem setup. 
	     (b) Temperature distribution under uniform thermal conductivity.}
\label{problem6}
\end{figure}

Problem 6 pertains to the utilization of smaller, non-identical 
heat sinks. As illustrated in Fig. \ref{problem6_set_up}, 
two heat sinks, each occupying $10\%$ of the side length,
are positioned at the center of the top (350K) and the 
bottom (300K) boundaries.
Reference solutions obtained by the AA and TGH methods 
are available in Ref. \cite{tong2018optimizing}.
The temperature distribution under uniform thermal conductivity 
as shown in Fig. \ref{problem6_non_tem}, is consistent with the 
reference results (their Fig. 6(d))
and the present average temperature (365.1K) 
matches that of the reference (363.3K) also. 

\begin{figure}[htb!]
	\centering
	\makebox[\textwidth][c]{\subfigure[]{
			\includegraphics[width=.45\textwidth]{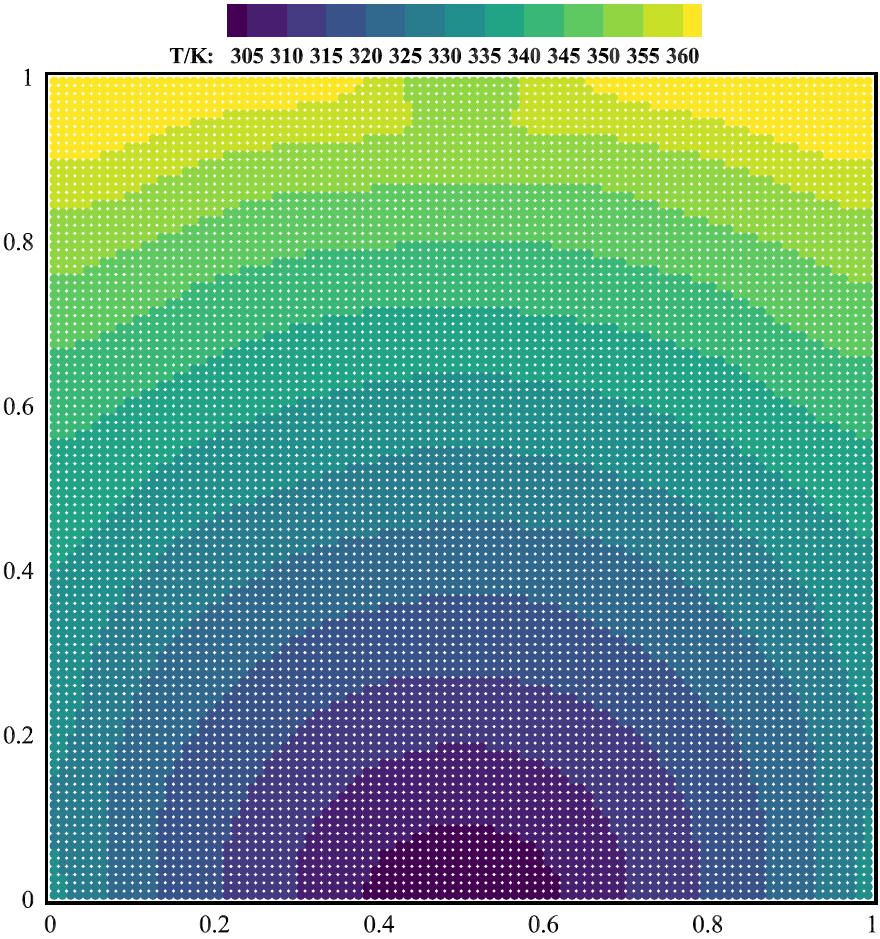}
			\label{problem6_op_tem_td}
		}
		\quad
		\subfigure[]{
			\includegraphics[width=.45\textwidth]{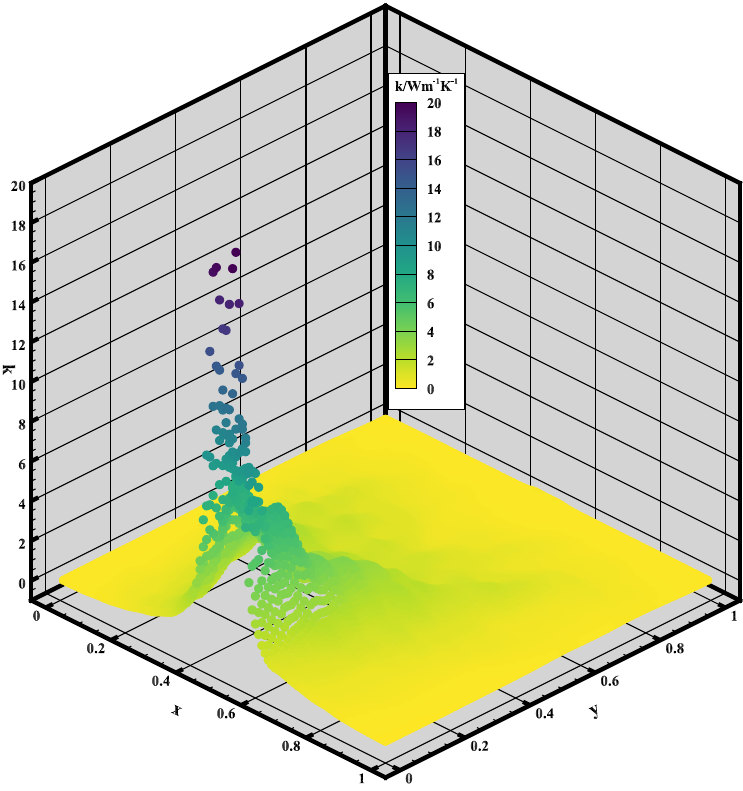}
			\label{problem6_op_k_td}
		}
		\quad}
	\caption{Present optimized results of the Problem 6.
		(a) Temperature;
		(b) Thermal conductivity.}
	\label{problem6_op_tem}
\end{figure}

The present optimization results are depicted in Fig. \ref{problem6_op_tem}, 
and Table \ref{problem6_comparison_temperature} provides a summary of comparisons.
It is observed that the present temperature reduction ratio 
is higher than that achieved by the TGH method,
although slightly lower than the result obtained using the AA method.
As shown in Fig. \ref{problem6_op_tem_td}, 
the present optimized temperature distribution 
presents a smoother profile compared to that of the reference (their Fig. 6(b)),
especially in the vicinity of the high-temperature sink. 
From the present thermal conductivity distribution
in Fig. \ref{problem6_op_tem_td}, 
the most notable feature is the single peak,
in contrast to four peaks obtained in the reference result (their Fig. 7(b)),
which refines the mesh around the small heat sinks.
Note that, due to the introduction of regularization,
the present highest value of thermal conductivity is only about 20W/(m·K),
which is significantly lower than the reference result of about 200W/(m·K).
Moreover, the computation time for a steady solution is 137.9 seconds, 
while the time required for the entire optimization is 890.8 seconds,
again showing quite good efficiency.
\begin{table}[htb!]
	\small
	\renewcommand\arraystretch{1.25}
	\centering
	\captionsetup{font={small}}
	\caption{Summary of results comparison for Problem 6.}
	\begin{tabularx}{14cm}{@{\extracolsep{\fill}}ccccc}
		\hline
		\quad Problem &Method & Original $\overline{T}\left(K\right)$ 
		& Optimized $\overline{T}\left(K\right)$ 
		& Reduced $\left(\% \right)$ \quad\\
		\midrule
		\quad \multirow{3}{*}{6} 
		&TD                            & $365.1$ & $333.0$ & $8.79$ \quad\\
		&AA \cite{tong2018optimizing}  & $363.3$ & $331.0$ & $8.89$ \quad\\
		&TGH \cite{tong2018optimizing} & $363.3$ & $331.9$ & $8.64$ \quad\\
		\bottomrule
	\end{tabularx}
	\label{problem6_comparison_temperature}
\end{table}

\subsection{The 1/10 sinks with the heat flux heater}
\begin{figure}
	\makebox[\textwidth][c]{\subfigure[]{
			\includegraphics[width=.45\textwidth]{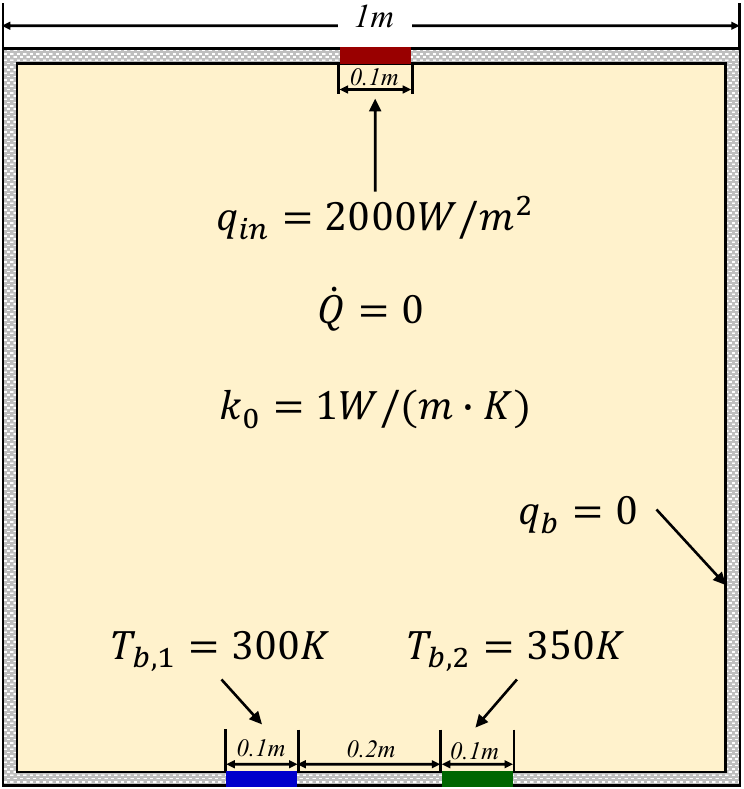}
			\label{problem7_set_up}
		}
		\quad
		\subfigure[]{
			\includegraphics[width=.45\textwidth]{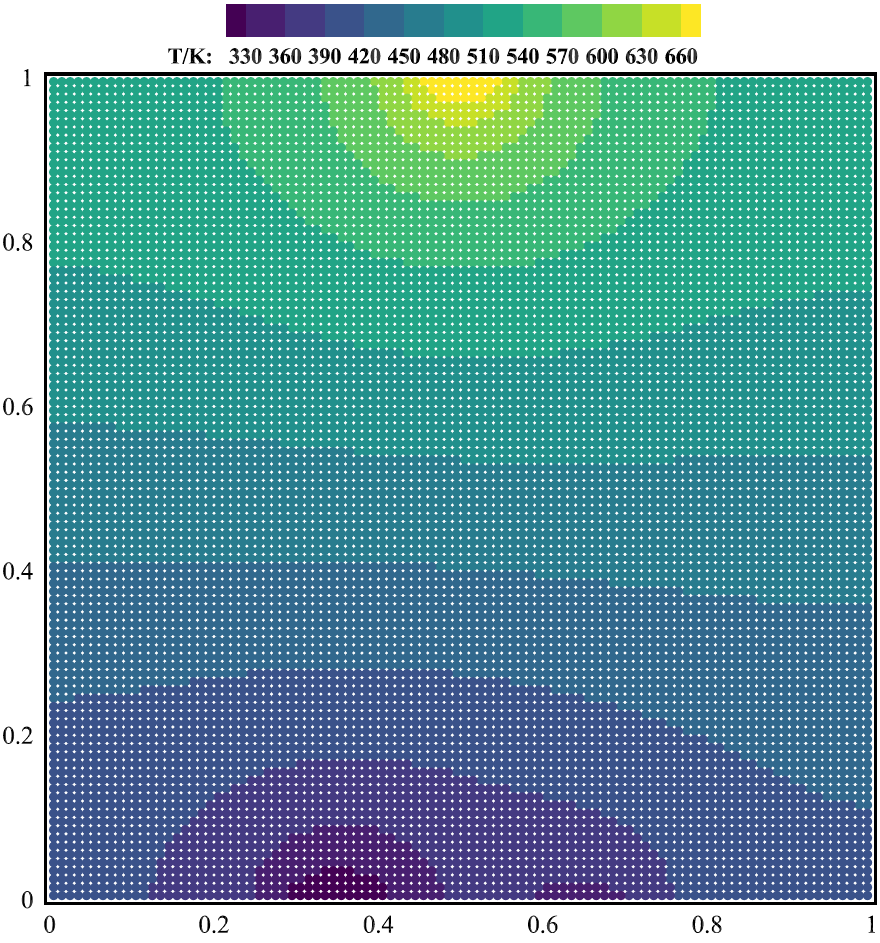}
			\label{problem7_non_tem}
		}
		\quad}
	\caption{Problem 7. 
		(a) Schematic depiction of the problem setup. 
		(b) Temperature distribution under uniform thermal conductivity.}
	\label{problem7}
\end{figure}

Problem 7 explores a thermal domain that features two heat sinks 
on one side with different operating temperatures, 
along with a heat flux heater situated on the opposite side.
As depicted in Fig. \ref{problem7_set_up}.
a heat flux heater with a magnitude of 
$\rm 2000W/m^{2}$ is positioned at the central of the upper boundary, 
covering $10\%$ of the side length,
and two heat sinks maintained at 300K and 350K, respectively, 
with the same size are located along the lower boundary.
Reference solutions with the target of minimizing temperature 
on the flux boundary can be found in Ref. \cite{tong2018optimizing}.
The temperature distribution with uniform thermal conductivity 
is depicted in Fig. \ref{problem7_non_tem},
and shows good agreement with the reference (their Fig. 10(d)), 
although the average temperature on the flux boundary is 
slightly lower than that in the reference, 
possibly due to the mesh refinement near
the flux heater in the latter.

\begin{figure}[htbp]
	\centering
	\makebox[\textwidth][c]{\subfigure[]{
			\includegraphics[width=.45\textwidth]{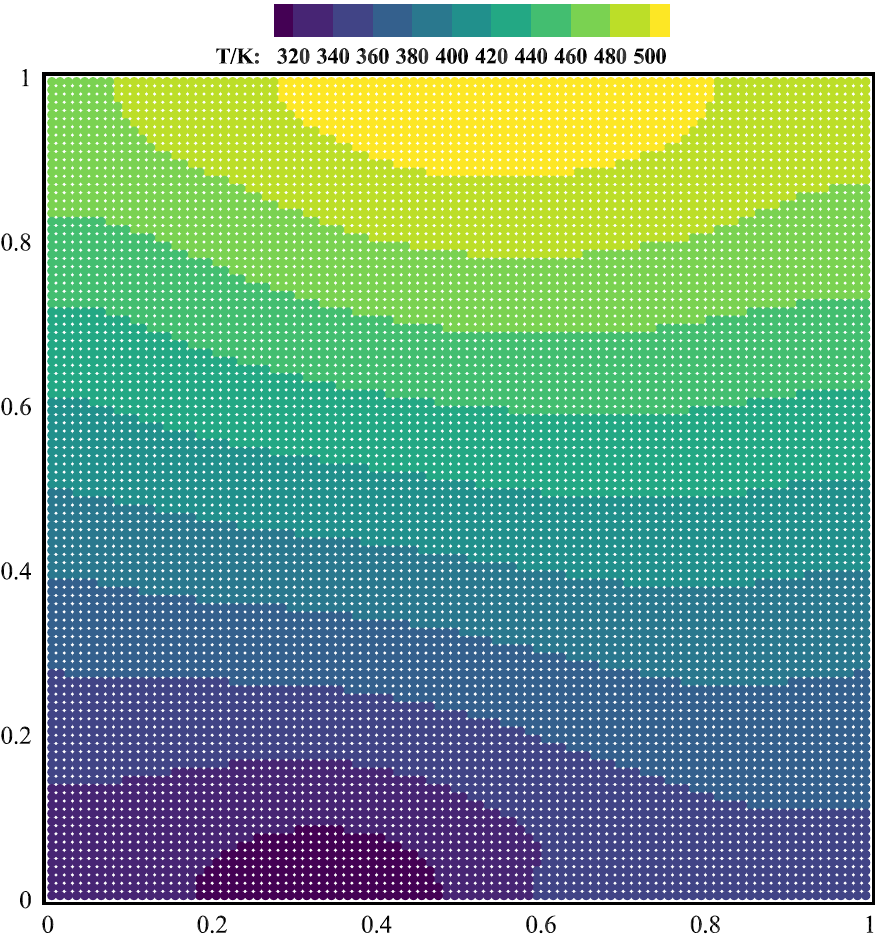}
			\label{problem7_op_tem_td}
		}
		\quad
		\subfigure[]{
			\includegraphics[width=.45\textwidth]{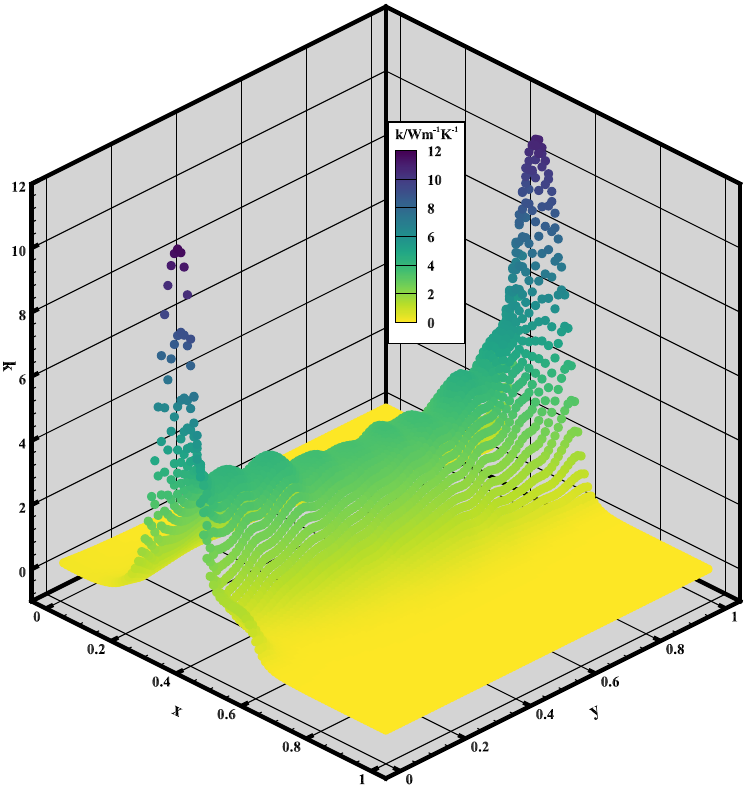}
			\label{problem7_op_k_td}
		}
		\quad}
	\caption{Present optimized results of the Problem 7.
		(a) Temperature; (b) Thermal conductivity.}
	\label{problem7_op_tem}
\end{figure}

The present results, with the objective of minimizing 
the average temperature of the domain, 
are shown in Fig. \ref{problem7_op_tem},
and the comparison of results is summarized in Table \ref{problem7_comparison}.
The optimized thermal conductivity contributes to an 
overall reduction in temperature.
Note that, even though the present objectives 
is different from that of the reference,
the pattern of the optimized thermal conductivity in 
Fig. \ref{problem7_op_k_td} closely resembles that of the reference 
(their Fig. 11(b)).
It forms a bridge with high thermal conductivity between the 
flux heater and the colder sink.
Compared to the reference results (their Fig. 10(b)),
the present optimized temperature distribution
also exhibits a uniform gradient perpendicular to the 
line connecting the heater and the colder sink, proving the 
enhanced cooling capacity of the colder sink. 
\begin{table}[htbp]
	\small
	\renewcommand\arraystretch{1.25}
	\centering
	\captionsetup{font={small}}
	\caption{Summary of results comparison for Problem 7.}
	\begin{threeparttable}
		\begin{tabularx}{14cm}{@{\extracolsep{\fill}}ccccc}
			\hline
			\quad Problem &Method & Original $\overline{T}\left(K\right)$ 
			& Optimized $\overline{T}\left(K\right)$ 
			& Reduced $\left(\% \right)$ \quad\\
			\midrule
			\quad \multirow{3}{*}{7}
			&TD\tnote{1}                    & $470.7$ & $415.3$ & $11.77$ \quad\\
			&TD\tnote{2}                    & $689.5$ & $519.5$ & $24.66$ \quad\\
			&AA \cite{tong2018optimizing}   & $693.2$ & $501.6$ & $27.64$ \quad\\
			\bottomrule
		\end{tabularx}
		\begin{tablenotes}
			\footnotesize
			\item[1] The average temperature across the entire thermal domain.
			\item[2] The average temperature along the heat flux boundary.
		\end{tablenotes}
	\end{threeparttable}
	\label{problem7_comparison}
\end{table}
However, due to the different objectives, 
the present result shows a higher averaged temperature 
on the flux boundary compared to the reference.
Again, the computation time for the steady solution is 197.9 seconds, 
whereas the optimization requires 970.8,
suggesting good efficiency.
%%%%%%%%%%%%%%%%%%%%%%%%%%%%%%%%%%%%%%%%%%%%%%%%%%%%%%%%%%%%%
%
% Section
%
%%%%%%%%%%%%%%%%%%%%%%%%%%%%%%%%%%%%%%%%%%%%%%%%%%%%%%%%%%%%%
\section{Conclusion and remark}\label{conclusion}
%%%%%%%%%%%%%%%%%%%%%%%%%%%%%%%%%%%%%%%%%%%%%%%%%%%%%%%%%%%%%
In this paper, a target-driven all-at-once approach  
for PDE-constrained optimization is introduced, 
and it is applied to optimize thermal conduction problems.
By splitting the optimization iteration 
into small, easily managed steps and 
treating both state and design variables in the same way, 
the need for deriving complex adjoint equations 
and obtaining converged state solutions 
at each optimization iteration is eliminated.
In addition, the mesh-free, splitting-operator based implicit SPH method 
is employed as the underlying numerical method,
and a diffusion-analogy regularization approach 
is developed to ensure the numerical stability.
Typical examples of thermal conduction problems
demonstrate that the present method is able to achieve 
quite efficient optimization 
with the computational cost generally 
on the same order as obtaining a single converged PDE solution.
Furthermore, the present optimal results are 
comparable to those from the previous work,
but with the industrially relevant advantage of lower extreme values. 
Note that, as the target-driven concept used in the 
present method is not restricted to specific optimization targets,
it may be extended for, such as those on the domain boundary, 
topology, or other thermal and fluid dynamics applications,
which are also our future work focus. 
%
%%%%%%%%%%%%%%%%%%%%%%%%%%%%%%%%%%%%%%%%%%%%%%%%%%%%%%%%%%%%%
%
% Section
%
%%%%%%%%%%%%%%%%%%%%%%%%%%%%%%%%%%%%%%%%%%%%%%%%%%%%%%%%%%%%%
\section*{Acknowledgments}
\addcontentsline{toc}{section}{Acknowledgement}
The first author would like to acknowledge the financial support provided by
the China Scholarship Council (No.202006230071).
C. Zhang and X.Y. Hu would like to express their gratitude to Deutsche
Forschungsgemeinschaft(DFG) for their sponsorship of this research under
grant number DFG HU1527/12-4.
The corresponding code of this work is available on GitHub at 
\url{https://github.com/Xiangyu-Hu/SPHinXsys}.

%\section*{References}
\bibliographystyle{elsarticle-num}
\bibliography{thermal_optimization}

\begin{thebibliography}{10}
\expandafter\ifx\csname url\endcsname\relax
  \def\url#1{\texttt{#1}}\fi
\expandafter\ifx\csname urlprefix\endcsname\relax\def\urlprefix{URL }\fi
\expandafter\ifx\csname href\endcsname\relax
  \def\href#1#2{#2} \def\path#1{#1}\fi

\bibitem{moore2014emerging}
A.~L. Moore, L.~Shi, Emerging challenges and materials for thermal management
  of electronics, Materials today 17~(4) (2014) 163--174.

\bibitem{almogbel1999conduction}
M.~Almogbel, A.~Bejan, Conduction trees with spacings at the tips,
  International Journal of Heat and Mass Transfer 42~(20) (1999) 3739--3756.

\bibitem{guo2003least}
Z.~Guo, X.~Cheng, Z.~Xia, Least dissipation principle of heat transport
  potential capacity and its application in heat conduction optimization,
  Chinese science bulletin 48 (2003) 406--410.

\bibitem{da2005distribution}
A.~Da~Silva, G.~Lorenzini, A.~Bejan, Distribution of heat sources in vertical
  open channels with natural convection, International Journal of Heat and Mass
  Transfer 48~(8) (2005) 1462--1469.

\bibitem{chen2016optimization}
K.~Chen, S.~Wang, M.~Song, Optimization of heat source distribution for
  two-dimensional heat conduction using bionic method, International Journal of
  Heat and Mass Transfer 93 (2016) 108--117.

\bibitem{feng2015constructalflowpassage}
H.~Feng, L.~Chen, Z.~Xie, F.~Sun, Constructal design for a disc-shaped area
  based on minimum flow time of a flow system, International Journal of Heat
  and Mass Transfer 84 (2015) 433--439.

\bibitem{ghani2017hydrothermal}
I.~A. Ghani, N.~A.~C. Sidik, N.~Kamaruzaman, Hydrothermal performance of
  microchannel heat sink: The effect of channel design, International Journal
  of Heat and Mass Transfer 107 (2017) 21--44.

\bibitem{dede2015topology}
E.~M. Dede, S.~N. Joshi, F.~Zhou, Topology optimization, additive layer
  manufacturing, and experimental testing of an air-cooled heat sink, Journal
  of Mechanical Design 137~(11) (2015).

\bibitem{zegard2016bridging}
T.~Zegard, G.~H. Paulino, Bridging topology optimization and additive
  manufacturing, Structural and Multidisciplinary Optimization 53 (2016)
  175--192.

\bibitem{bejan1997constructal}
A.~Bejan, Constructal-theory network of conducting paths for cooling a heat
  generating volume, International Journal of Heat and Mass Transfer 40~(4)
  (1997) 799--816.

\bibitem{de2015numerical}
J.~De~los Reyes, Numerical pde-constrained optimization. springerbriefs in
  optimization, Springer, Cham. doi 10 (2015) 978--3.

\bibitem{herzog2010lectures}
R.~Herzog, Lectures notes algorithms and preconditioning in pde-constrained
  optimization, Technical Report (2010).

\bibitem{herzog2010algorithms}
R.~Herzog, K.~Kunisch, Algorithms for pde-constrained optimization,
  GAMM-Mitteilungen 33~(2) (2010) 163--176.

\bibitem{ito2008lagrange}
K.~Ito, K.~Kunisch, Lagrange multiplier approach to variational problems and
  applications, SIAM, 2008.

\bibitem{zhao2022optimal}
T.~Zhao, X.~Wu, Z.-Y. Guo, Optimal thermal conductivity design for the
  volume-to-point heat conduction problem based on adjoint analysis, Case
  Studies in Thermal Engineering 40 (2022) 102471.

\bibitem{rall1996introduction}
L.~B. Rall, G.~F. Corliss, An introduction to automatic differentiation,
  Computational Differentiation: Techniques, Applications, and Tools 89 (1996)
  1--18.

\bibitem{abiodun2018state}
O.~I. Abiodun, A.~Jantan, A.~E. Omolara, K.~V. Dada, N.~A. Mohamed, H.~Arshad,
  State-of-the-art in artificial neural network applications: A survey, Heliyon
  4~(11) (2018).

\bibitem{song2021optimization}
M.~Song, K.~Chen, X.~Zhang, Optimization of the volume-to-point heat conduction
  problem with automatic differentiation based approach, International Journal
  of Heat and Mass Transfer 177 (2021) 121552.

\bibitem{chen2011alternative}
Q.~Chen, H.~Zhu, N.~Pan, Z.-Y. Guo, An alternative criterion in heat transfer
  optimization, Proceedings of the Royal Society A: Mathematical, Physical and
  Engineering Sciences 467~(2128) (2011) 1012--1028.

\bibitem{qi2015assessment}
W.~Qi, K.~Guo, H.~Liu, B.~Liu, C.~Liu, Assessment of two different optimization
  principles applied in heat conduction, Science Bulletin 60~(23) (2015)
  2041--2053.

\bibitem{chen2019entropy}
X.~Chen, T.~Zhao, M.-Q. Zhang, Q.~Chen, Entropy and entransy in convective heat
  transfer optimization: A review and perspective, International Journal of
  Heat and Mass Transfer 137 (2019) 1191--1220.

\bibitem{bertsekas2014constrained}
D.~P. Bertsekas, Constrained optimization and Lagrange multiplier methods,
  Academic press, 2014.

\bibitem{alt1993lagrange}
W.~Alt, K.~Malanowski, The lagrange-newton method for nonlinear optimal control
  problems, Computational Optimization and Applications 2 (1993) 77--100.

\bibitem{cheng2003constructs}
X.~Cheng, Z.~Li, Z.~Guo, Constructs of highly effective heat transport paths by
  bionic optimization, Science in China Series E: Technological Sciences 46
  (2003) 296--302.

\bibitem{cheng2009homogenization}
X.~Cheng, X.~Xu, X.~Liang, Homogenization of temperature field and temperature
  gradient field, Science in China Series E: Technological Sciences 52 (2009)
  2937--2942.

\bibitem{guo2007entransy}
Z.-Y. Guo, H.-Y. Zhu, X.-G. Liang, Entransy—a physical quantity describing
  heat transfer ability, International Journal of Heat and Mass Transfer
  50~(13-14) (2007) 2545--2556.

\bibitem{bejan1983entropy}
A.~Bejan, J.~Kestin, Entropy generation through heat and fluid flow (1983).

\bibitem{bejan1996entropy}
A.~Bejan, Entropy generation minimization: The new thermodynamics of
  finite-size devices and finite-time processes, Journal of Applied Physics
  79~(3) (1996) 1191--1218.

\bibitem{du2015optimization}
W.~Du, P.~Wang, L.~Song, L.~Cheng, Optimization of volume to point conduction
  problem based on a novel thermal conductivity discretization algorithm,
  Chinese Journal of Chemical Engineering 23~(7) (2015) 1161--1168.

\bibitem{wang2020constructal}
R.~Wang, Z.~Xie, Y.~Yin, L.~Chen, Constructal design of elliptical cylinders
  with heat generating for entropy generation minimization, Entropy 22~(6)
  (2020) 651.

\bibitem{chen2013entransyreview}
Q.~Chen, X.-G. Liang, Z.-Y. Guo, Entransy theory for the optimization of heat
  transfer--a review and update, International Journal of Heat and Mass
  Transfer 63 (2013) 65--81.

\bibitem{zhao2022irreversibility}
T.~Zhao, Y.-C. Hua, Z.-Y. Guo, Irreversibility evaluation for transport
  processes revisited, International Journal of Heat and Mass Transfer 189
  (2022) 122699.

\bibitem{zhao2019collaborative}
T.~Zhao, D.~Liu, Q.~Chen, A collaborative optimization method for heat transfer
  systems based on the heat current method and entransy dissipation extremum
  principle, Applied Thermal Engineering 146 (2019) 635--647.

\bibitem{wu2022study}
S.~Wu, K.~Zhang, G.~Song, J.~Zhu, B.~Yao, F.~Li, Study on the performance of a
  miniscale channel heat sink with y-shaped unit channels based on entransy
  analysis, Applied Thermal Engineering 209 (2022) 118295.

\bibitem{tong2018optimizing}
Z.-X. Tong, M.-J. Li, J.-J. Yan, W.-Q. Tao, Optimizing thermal conductivity
  distribution for heat conduction problems with different optimization
  objectives, International Journal of Heat and Mass Transfer 119 (2018)
  343--354.

\bibitem{zhang2023effective}
J.~Zhang, X.~Wu, M.~Song, K.~Chen, An effective method for hot spot temperature
  optimization in heat conduction problem, Applied Thermal Engineering (2023)
  120325.

\bibitem{xia2002heat}
Z.-Z. Xia, Z.-X. Li, Z.~Guo, Heat conduction optimization: high conductivity
  constructs based on the principle of biological evolution, in: International
  Heat Transfer Conference Digital Library, Begel House Inc., 2002.

\bibitem{xia2004bionic}
Z.-Z. Xia, X.-G. Cheng, Z.-X. Li, Z.-Y. Guo, Bionic optimization of heat
  transport paths for heat conduction problems, Journal of Enhanced Heat
  Transfer 11~(2) (2004).

\bibitem{boichot2009tree}
R.~Boichot, L.~Luo, Y.~Fan, Tree-network structure generation for heat
  conduction by cellular automaton, Energy Conversion and Management 50~(2)
  (2009) 376--386.

\bibitem{boichot2010simple}
R.~Boichot, L.~Luo, A simple cellular automaton algorithm to optimise heat
  transfer in complex configurations, International Journal of Exergy 7~(1)
  (2010) 51--64.

\bibitem{xu2007optimization}
X.~Xu, X.~Liang, J.~Ren, Optimization of heat conduction using combinatorial
  optimization algorithms, International journal of heat and mass transfer
  50~(9-10) (2007) 1675--1682.

\bibitem{burger2013three}
F.~H. Burger, J.~Dirker, J.~P. Meyer, Three-dimensional conductive heat
  transfer topology optimisation in a cubic domain for the volume-to-surface
  problem, International Journal of Heat and Mass Transfer 67 (2013) 214--224.

\bibitem{manuel2017design}
M.~C.~E. Manuel, P.~T. Lin, Design explorations of heat conductive pathways,
  International Journal of Heat and Mass Transfer 104 (2017) 835--851.

\bibitem{lucy1977numerical}
L.~B. Lucy, A numerical approach to the testing of the fission hypothesis, The
  astronomical journal 82 (1977) 1013--1024.

\bibitem{gingold1977smoothed}
R.~A. Gingold, J.~J. Monaghan, Smoothed particle hydrodynamics: theory and
  application to non-spherical stars, Monthly notices of the royal astronomical
  society 181~(3) (1977) 375--389.

\bibitem{vishwakarma2011steady}
V.~Vishwakarma, A.~K. Das, P.~Das, Steady state conduction through 2d irregular
  bodies by smoothed particle hydrodynamics, International Journal of Heat and
  Mass Transfer 54~(1-3) (2011) 314--325.

\bibitem{xiao2017modeling}
Y.~Xiao, H.~Zhan, Y.~Gu, Q.~Li, Modeling heat transfer during friction stir
  welding using a meshless particle method, International journal of Heat and
  mass Transfer 104 (2017) 288--300.

\bibitem{adami2012generalized}
S.~Adami, X.~Y. Hu, N.~A. Adams, A generalized wall boundary condition for
  smoothed particle hydrodynamics, Journal of Computational Physics 231~(21)
  (2012) 7057--7075.

\bibitem{ryan2010novel}
E.~M. Ryan, A.~M. Tartakovsky, C.~Amon, A novel method for modeling neumann and
  robin boundary conditions in smoothed particle hydrodynamics, Computer
  Physics Communications 181~(12) (2010) 2008--2023.

\bibitem{strang1968construction}
G.~Strang, On the construction and comparison of difference schemes, SIAM
  journal on numerical analysis 5~(3) (1968) 506--517.

\bibitem{nguyen2009mass}
K.~Nguyen, A.~Caboussat, D.~Dabdub, Mass conservative, positive definite
  integrator for atmospheric chemical dynamics, Atmospheric Environment 43~(40)
  (2009) 6287--6295.

\bibitem{bishop2006pattern}
C.~M. Bishop, N.~M. Nasrabadi, Pattern recognition and machine learning,
  Vol.~4, Springer, 2006.

\bibitem{zhu2022dynamic}
Y.~Zhu, C.~Zhang, X.~Hu, A dynamic relaxation method with operator splitting
  and random-choice strategy for sph, Journal of Computational Physics 458
  (2022) 111105.

\bibitem{bagnara1995unified}
R.~Bagnara, A unified proof for the convergence of jacobi and gauss--seidel
  methods, SIAM review 37~(1) (1995) 93--97.

\bibitem{alexandersen2021revisiting}
J.~Alexandersen, O.~Sigmund, Revisiting the optimal thickness profile of
  cooling fins: A one-dimensional analytical study using optimality conditions,
  in: 2021 20th IEEE Intersociety Conference on Thermal and Thermomechanical
  Phenomena in Electronic Systems (iTherm), IEEE, 2021, pp. 24--30.

\end{thebibliography}

\end{document}